\documentclass[11pt]{article} % use larger type; default would be 10pt

\usepackage[utf8]{inputenc} % set input encoding (not needed with XeLaTeX)
\usepackage{amsfonts}
\usepackage{booktabs}
\usepackage{siunitx}
\usepackage[margin=1.2in]{geometry}% to change the page dimensions
\usepackage{setspace}
\usepackage{kotex}
\usepackage{graphicx,color} % support the \includegraphics command and options
\usepackage{amsmath,amssymb}
\usepackage{amsthm}
\usepackage{hyperref} 
\usepackage[export]{adjustbox}
\usepackage{footnote,xcolor}
\hypersetup{linkbordercolor=green}
\usepackage{booktabs} % for much better looking tables
\usepackage{array} % for better arrays (eg matrices) in maths
\usepackage{paralist} % very flexible & customisable lists (eg. enumerate/itemize, etc.)
\usepackage{verbatim} % adds environment for commenting out blocks of text & for better verbatim
\usepackage{subfig} % make it possible to include more than one captioned figure/table in a single float
\usepackage{fancyhdr} % This should be set AFTER setting up the page geometry
\usepackage{sectsty}
\allsectionsfont{\sffamily\mdseries\upshape} % (See the fntguide.pdf for font help)
\usepackage[round]{natbib}   % omit 'round' option if you prefer square brackets

\usepackage[titles,subfigure]{tocloft} % Alter the style of the Table of Contents

\newcommand{\E}{\mathbf{E}}

\newcommand{\1}{\mathbf{1}}

\newtheorem{assumption}{Assumption}

\newtheorem{theorem}{Theorem}
\newtheorem{lemma}{Lemma}

 \usepackage{titlesec}
 
\usepackage{mdframed}

\titleformat*{\section}{\Large\bfseries}
\titleformat*{\subsection}{\large\bfseries}
\titleformat*{\subsubsection}{\large\bfseries}
\titleformat*{\paragraph}{\small\bfseries}
\titleformat*{\subparagraph}{\small}
\usepackage{enumitem}

\usepackage{enumitem}

\newlist{thmlist}{enumerate}{1}
\setlist[thmlist]{label=(\roman{thmlisti}),noitemsep}

\usepackage{thmtools}

\usepackage{hyperref}
\hypersetup{colorlinks=true, linkcolor=blue, filecolor=blue, urlcolor=blue, citecolor=blue} % use for hypertext
\usepackage[colorinlistoftodos]{todonotes}

         \renewenvironment{abstract}
 {\small
  \begin{center}
  \bfseries \abstractname\vspace{-.5em}\vspace{0pt}
  \end{center}
  \list{}{%
    \setlength{\leftmargin}{5mm}% <---------- CHANGE HERE
    \setlength{\rightmargin}{\leftmargin}%
  }%
  \item\relax}
 {\endlist}
\begin{document}
%\title{Treatment effect analysis under social interactions}
%\maketitle
%\tableofcontents
\title{Analysis of Randomized Experiments\\ with Network Interference and Noncompliance}
\author{Bora Kim}
\maketitle
\pagenumbering{roman}
%\tableofcontents

\begin{abstract}
Randomized experiments have become a standard tool in economics. In analyzing randomized experiments, the traditional approach has been based on the Stable Unit Treatment Value (SUTVA: \cite{rubin}) assumption which dictates that there is no interference between individuals. However, the SUTVA assumption fails to hold in many applications due to social interaction, general equilibrium, and/or externality effects. While much progress has been made in relaxing the SUTVA assumption, most of this literature has only considered a setting with perfect compliance to treatment assignment. In practice, however, noncompliance occurs frequently where the actual treatment receipt is different from the assignment to the treatment. In this paper, we study causal effects in randomized experiments with network interference and noncompliance. Spillovers are allowed to occur at both treatment choice stage and outcome realization stage. In particular, we explicitly model treatment choices of agents as a binary game of incomplete information where resulting equilibrium treatment choice probabilities affect outcomes of interest. Outcomes are further characterized by a random coefficient model to allow for general unobserved heterogeneity in the causal effects. After defining our causal parameters of interest, we propose a simple control function estimator and derive its asymptotic properties under large-network asymptotics. We apply our methods to the randomized subsidy program of \cite{dupas} where we find evidence of spillover effects on both short-run and long-run adoption of insecticide-treated bed nets. Finally, we illustrate the usefulness of our methods by analyzing the impact of counterfactual subsidy policies.
\end{abstract}
\textbf{Keywords}: causal inference, interference, spillover, networks, games of incomplete information, control function
\pagenumbering{arabic}
%=============================================================================%
% note
% compliance: one-sided?
% compliance type도 써보는 건 어떨까
% dupas의 research question은 SR subsidy가 LR adoption에 미치는 효과.. SR adoption 자체와는 다르겠지.
% 어쩌면 CRCT 같은 방법론이 더 나을지도 몰라
%=============================================================================%
\section{Introduction}
Randomized experiments have become a standard tool for causal inference in economics. In analyzing randomized experiments, the traditional approach is based on the Stable Unit Treatment Value (SUTVA: \cite{rubin}) assumption which dictates that there is no interference between individuals. However, there are many settings where the SUTVA assumption fails to hold. For instance, deworming treatment given to some student may affect academic achievements of other students through externality effects (See for instance, \cite{KM}). In labor market, \cite{crepon} show that a large-scale job placement program affects non-participant’s employment probability through general equilibrium effects. \cite{ferracci} also report similar results. In such cases, there is interference or spillover effect where an individual’s behavior either directly or indirectly affects others’ outcomes through social interactions, externalities, or general equilibrium effects.

In recent years, there has been substantial progress in relaxing the SUTVA assumption in causal inference framework. Examples include \cite{manski}, \cite{HH}, \cite{L19}, \cite{VB20}, and \cite{baird}. 
Much of the literature, however, has been built on the restrictive assumption of perfect compliance to intervention in which experimental units perfectly comply with their assignment of treatment. In practice, noncompliance occurs commonly --- some units assigned to treatment group may opt out of the treatment, while some units assigned to control group may decide to take the treatment. In studies of labor market, for example, \cite{crepon} report that only 35$\%$ of those who were offered intensive job counseling actually took up the offer. While instrumental variables (IV) methods are widely used to address the noncompliance problem, these methods are developed based on the assumption that rules out interference between units (\cite{IA}).

The goal of this paper is to develop a formal framework to conduct causal inference in randomized experiments with \emph{both} spillovers and noncompliance. In the presence of noncompliance, spillovers can occur at two stages: at the treatment decision stage, and at the outcome realization stage. In the first stage in which each agent \emph{chooses} their treatment status, spillovers may occur if the utility from choosing treatment depends on the treatment choices of others. In the second stage where outcomes (or responses) are realized, agent's outcome can be affected not only by their own treatment choice, but also by treatment choices of others either directly or indirectly. While most of existing literature has only addressed the spillover effects at the outcome level (i.e., at the second stage), we allow for spillover effects both at the treatment choice (first stage) and at the outcome (second stage). 

To model spillovers, we take a game-theoretic approach. We consider a first stage model in which agents play a binary game of incomplete information. Such binary games of incomplete information have been used in various economic applications, e.g., in empirical industrial organization literature (\cite{bajari}), to model binary choices under peer effects (\cite{BD1}, \cite{BD2} and \cite{xu18}), and recently, to model network formation process (\cite{L14}, and \cite{RS}). We apply the method to the problem of endogenous treatment choices in the presence of spillovers. Specifically, we assume that agents simultaneously choose their treatment status as to maximize their expected utilities, given beliefs about anticipated treatment choices of their neighbors. In equilibrium, agents' subjective beliefs coincide with objective choice probabilities. Assuming that the unique equilibrium exists, the reduced-form model of agent's treatment choice can be written as a single threshold-crossing model where the threshold is a function of agent's own treatment assignment and the average equilibrium treatment choice probability of their neighbors. In the second stage, outcomes are modeled as being a function of agent's own treatment choice and the equilibrium average treatment choice probability of their neighbors, as it is determined in the first stage game. As in the first stage choice model, spillovers are captured by the equilibrium treatment choice probabilities. 

In our model, therefore, equilibrium treatment choice probabilities work as a mediator of spillover effects. This is different from the existing literature which often models the spillover at the outcome level by the proportion of treated neighbors. See for instance \cite{HH}, \cite{L19}, and \cite{VB20}. As we show later, when the outcome of interest represents a choice or behavior of individuals, their formulation implicitly assumes that the proportion of treated neighbors is fully observable to agents, i.e., agents possess a complete information over behaviors of their peers. However, the assumption of complete information is unrealistic especially in a single large network setting as ours where each individual has a considerable number of peers.\footnote{In our application, for instance, agents have 17 neighbors on average.} In such cases, it is more reasonable to assume that agents face uncertainty over others' behavior, making an incomplete information framework more adequate approximation of reality. 

We then characterize outcomes as a random coefficient model to allow for general unobserved heterogeneity. Our parameters of interest are average causal effects which include an average direct effect of own treatment take-up and an average spillover effect from direct neighbors. After rigorously defining our parameters of interest, we show our identification result. We first note that under general unobserved heterogeneity, the conventional instrumental variables (IV) methods do not identify the causal parameters when we allow for general heterogeneity in the outcome. We therefore propose our alternative identification based on a control function approach.

We then propose a simple two-step estimator where the first step estimates the payoff parameters of treatment choice games using nested fixed-point maximum-likelihood estimation and the second step estimates the average potential outcome functions using control function regression. Our estimator extends canonical \cite{heckman_sel} sample selection estimator (``Heckit”) to incorporate possible spillover effects. We show that the estimators are $\sqrt{n}$-consistent and asymptotically normal under the ``large-network” asymptotics in which a number of individuals connected in a single network increases to infinity. We study finite-sample properties of our estimators through Monte Carlo simulation.

Our methods are applied to the randomized subsidy program of \cite{dupas}. While the use of insecticide-treated nets (ITNs) has been shown to be effective in controlling malaria, the rate of adoption remains low. Given that the mosquito nets need to be re-purchased and replaced regularly, understanding the factors affecting household's short-run and long-run decision to purchase the bednet is an important task to achieve sufficiently high equilibrium adoption rate. In our application, we study the effect of short-run purchase of the bednet on the long-run purchase decision while incorporating possible spillovers from neighbors defined by geographical proximity. The treatment is a binary is a binary indicator for purchasing a mosquito net in the short-run (in Phase 1) and the outcome is a binary indicator for purchasing a mosquito net in the long-run (in Phase 2). 

We find evidence of positive spillover effects in the short-run bednet purchase decision. More specifically, in Phase 1, households were more likely to purchase the bednet when the average expected purchase rate of their neighbors is higher. On the contrary, we find the evidence of negative spillover effects in the long run although the statistical power is limited. Specifically, households were less likely to purchase the bednet in Phase 2 when the average expected purchase rate in Phase 1 was higher. Our results also suggest that the average direct effect of the bednet purchase in Phase 1 on the purchase in Phase 2 declines monotonically with respect to the expected neighborhood purchase rate in Phase 1. When the Phase-1 neighborhood purchase rate was $0\%$ (no spillover), households who purchased the bednet in Phase 1 were 36.9 percentage points more likely to purchase the bednet in Phase 2 compared to those who did not purchase the bednet in Phase 1. Such effect becomes almost to zero at another extreme where the neighborhood purchase rate was $100\%$ (full spillover). Ignoring spillover effects leads to the misleading conclusion that the average direct effect of the short-run purchase on the long-run purchase is almost zero when in fact, the effect varies from $0\%$ to $36\%$ depending on the degree of spillovers.

Our structural modeling allows researchers to analyze the impact of counterfactual policies on the outcome of interest. We illustrate this by analyzing the impact of counterfactual subsidy program on the long-run adoption in which a policy-maker implements a means-tested subsidy rule where the subsidy is given only when the household's income level is below some pre-specified threshold. We predict the average long-term adoption rate under different subsidy regimes defined by different values of the eligibility threshold. We find that even under the very generous subsidy regime where almost everyone in the sample receives the subsidy, the average long-run adoption rate does not exceed $20\%$, due to the large negative spillover in the long-run.

\paragraph{Related Literature}\ \\
Recent works on causal inference under spillovers mainly concentrate on the case with random treatment, i.e., they do not address treatment choice endogeneity. Examples include  \cite{HH}, \cite{L19}, and \cite{VB20}.  

In causal inference literature, game-theoretic models have been used in several papers. \cite{lazzati} proposes a structural model of treatment responses using games of complete information. However, the paper does not address the endogeneity of treatment choices.  \cite{han} allow spillovers at both choice and outcome stages using game theoretic approach. Their model is different from ours in that they model treatment choice by a binary game of complete (perfect) information. Also, \cite{han} consider an interaction within groups while we consider an interaction under general network. While the assumption of complete information may be appropriate under interactions in a relatively small group, incomplete information assumption is more reasonable under network interactions, especially when the network size is large. \cite{matt20} model treatment choices as a binary game of incomplete information. However, they do not consider spillovers at the outcome level while we are interested in separately identifying the individual treatment effect and spillover effect.

Meanwhile a literature from statistics has started to incorporate spillovers and noncompliance in network setting. See \cite{imai} for the most recent progress. Unlike our game-theoretic model, their model is reduced-form in nature and consequently, important aspects of economic mechanism behind treatment choices such as utility maximization are largely ignored. 

\paragraph{Outline}\ \\ We describe our model in Section \ref{sec:model}. We first outline our model of treatment choices and then the model of potential outcomes. Parameters of interest are also discussed. Section \ref{sec:idd} discusses identification of parameters of interest. We first show that the conventional IV methods are not valid in the presence of treatment effect heterogeneity. We then show how to use control function approach to achieve point identification. In Section \ref{sec:estimation}, we propose a simple two-stage estimation procedure. Asymptotic properties are derived and simulation results are also presented. Section \ref{sec:application} applies our methods to empirical setting.

%=============================================================================%
\section{Model of Treatment Choices and Outcomes}\label{sec:model}
%=============================================================================%
%---------------------------------choice model

In this section, we first describe our treatment choice models as a binary game under incomplete information. We then describe our model of treatment responses under spillovers. 

Let $\mathcal{N}_n=\{1,\cdots,n\}$ denote a set of agents. $n$-many agents are connected through a single, large network. Let $G$ be a symmetric $n\times n$ adjacency matrix where $ij$th entry ($G_{ij}$) represents a connection or link between agents. Specifically, $G_{ij}=1$ if agent $i$ and $j$ are connected and $G_{ij}=0$ otherwise. We assume $G_{ii}=0$ for all $i\in \mathcal{N}_n$ (no self-link). When $G_{ij}=1$, we say that $i$ and $j$ are (direct) \emph{peers} or \emph{neighbors}. Let $\mathcal{N}_i$ be a set of $i$'s peers, i.e., $\mathcal{N}_i=\{j\in\mathcal{N}_n:G_{ij}=1\}$. The number of $i$'s neighbors or \emph{degree} of $i$ is denoted as $|\mathcal{N}_i|$. 

%=============================================================================%
\subsection{Treatment Choice Model with Spillovers}\label{sec:model1}
%=============================================================================%
We consider a game theoretic model of treatment choice. Specifically, we characterize a realized treatment choice as a solution to a binary game under incomplete information played by agents in a given network. In this framework, agents simultaneously choose their treatment status in order to maximize their expected utility, given beliefs about the anticipated behaviors of their peers.

%-------------------------------------------------------------------------------------------------------------------------------------%
\paragraph{Utility}
Each agent $i$ has a vector of observed characteristics $X_i\in\mathcal{X}$ and an unobserved utility shock $v_i\in \mathbb{R}$. Throughout the paper, we assume that $\mathcal{X}$ is a bounded subset of $\mathbb{R}^{k}$.
 In addition, each $i$ is randomly assigned to treatment. Let $Z_i\in \{0,1\}$ represent $i$’s randomized treatment assignment where $Z_i=1$ if $i$ is assigned to treatment and $Z_i=0$ if $i$ is assigned to control. Let $Z=(Z_i)_{i\in\mathcal{N}_n}$ and $X=(X_i)_{i\in\mathcal{N}_n}$. 
  There is noncompliance if $Z\not=D$, i.e., for some $i$, the treatment assignment is different from the actual treatment received. There are two possible cases for this: $(Z_i,D_i)=(1,0)$ and $(Z_i,D_i)=(0,1)$. The former indicates that $i$ who was assigned to treatment group has refused to take the treatment. The latter indicates that  $i$ has received the treatment even when $i$ was assigned to control group. In this paper, we allow for both cases, i.e., we consider a setting with two-sided noncompliance. 

Unlike $Z_i$, $D_i$ is self-selection. We assume that each $i$ chooses $D_i\in\{0,1\}$ by utility maximization where the utility that $i$ receives depends on the choices of $i$’s peers. Let the utility function of agent $i$ be $\pi(D_i,D_{-i},X_i,Z_i,v_i)$ where $D_{-i}\in \{0,1\}^{n-1}$ is a vector of treatment choices of agents except for $i$. We specify the utility function as the following linear model:

\begin{eqnarray}
\pi(D_i,D_{-i},X_i,Z_i,v_i)
=\begin{cases}
X_i'\theta_{1}+\theta_2Z_i+\theta_3\frac{1}{|\mathcal{N}_i|}\sum_{j\in \mathcal{N}_i}D_j-v_{i}&\textrm{if }D_i=1\\
0 &\textrm{if } D_i=0.
\end{cases}
\end{eqnarray}

First note that the utility from choosing $D_i=0$ is normalized as zero. This is without loss of generality as only difference in utilities is identified. Utility of choosing $D_i=1$ depends on other agents' treatment choices through the term $\sum_{j\in \mathcal{N}_i}D_j/|\mathcal{N}_i|$, the fraction of peers taking up the treatment. This term represents social interactions or spillover effects in treatment choice. When $\theta_3=0$, there are no spillovers and the model becomes a usual single-agent binary choice model as in \cite{mcfadden}. When $\theta_3>0$, we have positive spillovers where the utility of choosing $D_i=1$ is higher when members of $i$’s reference group (directed neighbors in our specification) behave similarly. $\theta_3>0$ thus implies that agents have preference for conformity. On the other hand, when $\theta_3<0$, we conclude that there are negative spillovers in treatment choice.

%-------------------------------------------------------------------------------------------------------------------------------------%
% complication arises since i cannot observe… simultaneity… 따라서 expected utility. Information set specification은 이때쯤 튀어나오는 편이 좋겠네요… v_i와 S의 정의…
We assume that $v_i$ is a private information, i.e., $v_i$ is known only to $i$, and other agents cannot observe $v_i$. Therefore agents have incomplete information over others' choices. In other words, $i$ cannot observe other players' treatment choices at the time their choice is made. Instead, each agent $i$ chooses an action that maximizes their expected utility given their beliefs on $\sum_{j\in \mathcal{N}_i}D_j/|\mathcal{N}_i|$. Beliefs are formed under the information set available to $i$. Let $\tau_i$ denote $i$'s information set. We specify $\tau_i$ as follows: 
%-------------------------------------------------------------------------------------------------------------------------------------%

\begin{assumption}[informational structure]\label{a3} Let $G=(G_{ij})_{i,j\in\mathcal{N}_n}$, $X=(X_i)_{i\in \mathcal{N}_n}$ and $Z=(Z_i)_{i\in\mathcal{N}_n}$. We assume that $(G,X,Z)$ is a \emph{public information}, i.e., every agent knows the entire network structure ($G$), the vector of observed characteristics ($X$) and the vector of treatment assignment ($Z$). On the other hand, $v_{i}$ is a \emph{private information} of $i$ where its value is only known to $i$. Therefore $\tau_{i}=(G,X,Z,v_{i})$ summarizes the information available to $i$. 
\end{assumption}
The assumption \ref{a3} is standard in the literature on games of incomplete information. Let $S=(G,X,Z)$ be the set of public information. This is often called a \emph{public state variable} as well. For private information $v_i$, we make the following assumption:

%-------------------------------------------------------------------------------------------------------------------------------------%
\begin{assumption}[unobserved heterogeneity]\label{a4} For all $i\in\mathcal{N}_n$, a private information $v_i$ is 
\begin{enumerate}
\item[(i)] i.i.d. with a standard normal cdf $\Phi$ and 
\item[(ii)] independent of $S$.
\end{enumerate}
\end{assumption}
%-------------------------------------------------------------------------------------------------------------------------------------%
As in the standard single-agent binary choice models, distribution of $v_i$ must be known up to a finite-dimensional parameter. We use the normal distribution only for convenience. Other distributional assumptions such as logit can be used as well. The assumption that $v_i$'s are independent to each other is critical for our identification analysis. This assumption implies that the knowledge of $v_i$ does not help predicting $v_j$ for any $j\not=i$. To our knowledge, identification of incomplete information games with correlated private information in a general network setting is an open question.  Assumption \ref{a4} (ii) is trivially satisfied if we treat $S$ as fixed. Consequently, we do not address the issue of network endogeneity as it is not a focus of this paper.

%--------------------------------------------------------------------------------------------------------------------------------------%

%============================================================================%
\paragraph{Strategy} Let $D_{i}(\tau_i,\theta)$ denote $i$’s pure strategy which maps $i$’s information set $\tau_{i}=(S,v_{i})$ to a treatment choice $D_i\in \{0,1\}$ given a parameter value $\theta=(\theta_1,\theta_2,\theta_3)$. Agent $i$ chooses her optimal action by maximizing her expected utility $\E[\pi(D_i,D_{-i},X_i,Z_i,v_i)|\tau_i]$ where the expectation is taken with respect to $D_{-i}$ given her belief about $D_{-i}$. Let $\sigma_{j,i}$ be $i$'s belief over the event $\{D_{j}=1\}$ given the information $\tau_{i}$. Then 
\begin{eqnarray}
\sigma_{j,i}&=^{def}&\Pr(D_{j}=1|\tau_i)\\
&=&\Pr(D_{j}(\tau_j,\theta)=1|\tau_i)\\
&=&\Pr(D_j(S,v_j,\theta)=1|S,v_i)\\
&=&\Pr(D_{j}(S,v_j,\theta)=1)\\
&=&\sigma_j(S,\theta)
\end{eqnarray}
where the fourth equality follows from the Assumption \ref{a4}.  From the last equality, we see that $\sigma_{j,i}=\sigma_{j}$ for all $i\not=j$, i.e., every agent shares a common belief on $j$'s choice. This common belief should be consistent with actual probability of $j$ choosing $D_{j}=1$ under rational expectations as we show below.
%============================================================================%
\paragraph{Equilibrium}
Given the belief profile of $\{\sigma_j(S,\theta)\}_{j\not=i}$, agent $i$ calculates the expected utility he gets when choosing $D_i=1$ as follows:
\begin{eqnarray}
\E\big[\pi(1,D_{-i},X_i,Z_i,v_i)|\tau_{i}\big]&=&\E\big[X_i'\theta_1+\theta_2 Z_i+\theta_3\frac{1}{|\mathcal{N}_i|}\sum_{j\in \mathcal{N}_{i}}D_{j}-v_{i}\big|S,v_{i}\big]\\
&=&X_i'\theta_1+\theta_2 Z_i+\theta_3\frac{1}{|\mathcal{N}_i|}\sum_{j\in \mathcal{N}_{i}}\underbrace{\Pr(D_{j}=1|S)}_{= \sigma_{j}(S,\theta)}-v_{i}\\
&=&X_i'\theta_1+\theta_2 Z_i+\theta_3\frac{1}{|\mathcal{N}_i|}\sum_{j\in \mathcal{N}_{i}}\sigma_{j}(S,\theta)-v_{i}.
\end{eqnarray}
Agent $i$ would choose $D_{i}=1$ if $\E\big[\pi(1,D_{-i},X_i,Z_i,v_i)|\tau_{i}\big]\ge 0$. Therefore, 
$$D_{i}=\1\Big\{v_{i}\le X_i'\theta_1+\theta_2 Z_i+\theta_3\frac{1}{|\mathcal{N}_i|}\sum_{j\in \mathcal{N}_{i}}\sigma_{j}(S,\theta)\Big\}.$$
 Bayes-Nash equilibrium (BNE) is defined by a vector of choice probabilities $\sigma^*(S,\theta)=\big(\sigma_i^*(S,\theta)\big)_{i\in\mathcal{N}_n}$ that is consistent with the observed decision rule in the sense that it satisfies the following system of equations:
\begin{eqnarray}
\sigma_i^{*}(S,\theta)&=&\Pr\big(v_{i}\le X_i'\theta_1+\theta_2 Z_i+\theta_3\frac{1}{|\mathcal{N}_i|}\sum_{j\in \mathcal{N}_{i}}\sigma_{j}^{*}(S,\theta)\big),\quad \forall i\in\mathcal{N}_n\\
&=&\Phi\Big(X_i'\theta_1+\theta_2 Z_i+\theta_3\frac{1}{|\mathcal{N}_i|}\sum_{j\in\mathcal{N}_i}\sigma_j^{*}(S,\theta)\Big),\quad \forall i\in\mathcal{N}_n.\label{sss}
\end{eqnarray}
Here we use the superscript $[*]$ to emphasize that $\sigma^*(S,\theta)$ is an equilibrium quantity. In other words, Bayes-Nash equilibrium given $(S,\theta)$ is a vector $\sigma^*(S,\theta)$ which is defined as a fixed point to the system of equations above. By the implicit function theorem, it can be shown easily that $\sigma^*(S,\theta)$ is smooth in both $S$ and $\theta$. Therefore the existence of a fixed point is guaranteed due to Brouwer's fixed point theorem for any realized data $S$ and parameter value $\theta$. However, there can be many fixed points $\sigma^{*}(S,\theta)$ solving the system. We show that a unique equilibrium exists if we restrict the value of $\theta_3$ to be sufficiently mild. Formally,

%--------------------------------------------------------------------------------------------------------------------------------------%
\begin{theorem}[unique equilibrium]\label{ue} Let the pdf of $v_i$ be $\phi(v)$. Define $\lambda=|\theta_3|\sup_u\phi(u)$. For any $S$ and $\theta$, there exists a unique equilibrium $\{\sigma^{*}_j(S,\theta)\}_{j\in \mathcal{N}_n}$ if $\lambda<1$.
\end{theorem}
See appendix \ref{uniq} for proof. When $v_i$ is normally distributed, we have $\sup_u\phi(u)=1/\sqrt{2\pi}$. Therefore $\lambda<1$ is equivalent to $|\theta_3|<\sqrt{2\pi}\approx 2.5$. Throughout the paper we assume that $\lambda<1$ so that the degree of interaction is not too strong to breed multiple equilibria. 

\begin{assumption}[unique equilibrium] $|\theta_3|<\sqrt{2\pi}$. 
\end{assumption}

Under the unique equilibrium, agent's treatment choice can be written as the following reduced-form equation:
\begin{eqnarray}
&&D_{i}=\1\Big\{v_i\le X_i'\theta_1+\theta_2 Z_i+\theta_3\frac{1}{|\mathcal{N}_i|}\sum_{j\in \mathcal{N}_{i}}\sigma_{j}^*(S,\theta)\Big\}\label{Dp1}\\
&\Longleftrightarrow&D_{i}=\1\Big\{\Phi(v_i)\le \Phi\Big(X_i'\theta_1+\theta_2 Z_i+\theta_3\frac{1}{|\mathcal{N}_i|}\sum_{j\in \mathcal{N}_{i}}\sigma_{j}^*(S,\theta)\Big)\Big\}\label{Dp2}\\ 
&\Longleftrightarrow&D_{i}=\1\Big\{\Phi(v_i)\le \sigma_i^*(S,\theta)\Big\}\label{Dp3}
\end{eqnarray}
where the last step follows from \ref{sss}. 

The story goes like this: For given $S$ and $\theta$, the equilibrium choice probabilities $\sigma_i^*(S,\theta),\forall i\in\mathcal{N}_n$ are realized. Observing this equilibrium, each agent chooses their treatment status according to either \ref{Dp1},\ref{Dp2} or \ref{Dp3}.

%=============================================================================%%=============================================================================%
\subsection{Potential Outcomes Model with Spillovers}\label{sec:model2}
In this section, we propose our model of treatment response in settings with spillovers. Previous research on treatment response has been based on the SUTVA assumption which requires that an individual’s outcome depends only on their own treatment status. Under the SUTVA assumption, $i$’s outcome or response $Y_i$ can be written as $Y_i=Y_i(D_i)$. Let $d\in\{0,1\}$ be the possible treatment value that agents can get. Potential outcome under the SUTVA assumption is denoted by $Y_i(d)$, which delivers the response of $i$ when assigned to $D_i=d$. Unlike the SUTVA case, however, there is no obvious way to model spillovers in the treatment response. As \cite{manski} and \cite{KT20} show, there are many ways to relax the SUTVA assumption, each of which is based on different restrictions on the nature of interference between agents. 

In our paper, we assume that $i$'s outcome is a function of a \emph{direct effect} from own treatment status and an \emph{indirect effect} or \emph{spillover effect} from $i$'s neighbors. Spillover effects are assumed to be mediated by $\sum_{j\in\mathcal{N}_i}\sigma_j^*(S,\theta)/|\mathcal{N}_i|$. For notational simplicity, let us define $\pi_i^*(S,\theta)=\sum_{j\in\mathcal{N}_i}\sigma_j^*(S,\theta)/|\mathcal{N}_i|$. Also, $\pi_i^*$ and $\pi_i^*(S,\theta)$ will be used interchangeably. Thus, we write the realized outcome of $i$ as follows:

$$Y_i=Y_i(D_i,\pi_i^*) $$
where $\pi_i^*=\pi_i^*(S,\theta)$ is the average of equilibrium treatment choice probabilities of $i$'s neighbor. From now on, we simply refer to $\pi_i^*$ as $i$'s ``\emph{neighborhood (propensity) score}”. This is the average value of propensity scores of $i$'s direct neighbors where each score measures the probability of taking up the treatment given the public information $S$. \cite{matt20} have termed the same object as ``peer-influenced propensity score”.

Let $\pi\in[0,1]$ be the possible value that $\pi_i^*$ can take.  The potential outcome $Y_i(d,\pi)$ represents $i$'s response when we exogenously assign $D_i=d$ and $\pi_i^*=\pi$. Concretely, $Y_i(1,\pi)$ represents $i$'s outcome when $i$ is required to be treated and $i$'s neighborhood score has been exogenously set to $\pi$. Similarly $Y_i(0,\pi)$ is $i$'s outcome when $i$ is forbidden to be treated and $i$'s neighborhood score has been exogenously set to $\pi$. Underlying assumption is that it is possible to manipulate the value of $D_i$ and $\pi_i^*$. Since $\pi_i^*$ is a function of the public state variable $S=(G,X,Z)$, we can conceivably manipulate the value of $\pi_i^*$ by changing $Z$ for a given $(G,X)$, which is assumed to be predetermined and non-manipulable. Thus $Y_i(d,\pi)$ can be realized through changing $Z$ profile in the population in a way that it induces $\pi_i^*=\pi$ as an equilibrium in the first-stage and then requiring $i$ to choose $D_i=d$. \footnote{Note that some combination $(d,\pi)$  may represent off-the-equilibrium quantity. Thus, the resulting $Y_i(d,\pi)$ may not be a policy-relevant counterfactual. Nevertheless, to define causal effects rigorously, we need to consider every possible combinations of $(d,\pi)\in\{0,1\}\times [0,1]$.}

%--------------------------------------------------------------------------------------------------------------------------------------%

%\paragraph{Note} $\pi_i(Z_n)\approx D_i$. Define the difference as $\eta_i=D_i-\E[D_i|Z_n]$. By definition, $\E[\eta_i|Z_n,X_n,G_n]=0$ for any realization of conditioning variables.
%--------------------------------------------------------------------------------------------------------------------------------------%

\paragraph{Comparison to other approaches} The existing literature with interference often models potential outcomes as a function of own treatment status and the proportion of treated neighbors or the number of treated neighbors (e.g. \cite{HH}, \cite{L19}, \cite{VB20}).  Define $\bar D_i\equiv\sum_{j\in \mathcal{N}_i}D_j/|\mathcal{N}_i|$ with a generic value $\bar d\in[0,1]$. Such models then write the realized outcome as $Y_i=Y_i(D_i,\bar D_i)$ and the potential outcomes as $Y_i(d,\bar d)$. Our model differs from theirs in that we model spillovers via \emph{ex ante} (anticipated) expectation of $\bar D_i$ rather than \emph{ex post} realization of $\bar D_i$ itself. Recall that $\pi_i^*(S,\theta)=\E[\bar D_i|S]$. Since the difference between $\bar D_i$ and $\pi_i^*(S,\theta)$ has a mean zero (i.e., $\E[\bar D_i-\pi_i^*(S,\theta)|S]=0$), in practice the values of these two quantities may not be too different, especially when $|\mathcal{N}_i|$ is large.

Nevertheless, they are based on two different behavioral assumptions. Suppose that the outcome of interest represents decision or behavior of agents. Then the formulation $Y_i=Y_i(D_i,\bar D_i)$ is derived under the assumption that agents base their decisions on $\bar D_i$ rather than expected $\bar D_i$. This is realistic only when $\bar D_i$ is fully observed at the time decision on $Y_i$ is made. Thus, the model could be interpreted as a model with complete or perfect information. On the other hand, our specification $Y_i=Y_i(D_i,\pi_i^*)$ assumes that agents do not fully observe $\bar D_i$ when they decide their $Y_i$. Thus agents face an intrinsic uncertainty over others' treatment choices even at the second-stage. This is plausible when the reference group is relatively large so that it is not easy for agents to fully observe the value of $\bar D_i$. Also, there are settings where agents are reluctant to reveal their treatment status ---  For instance when treatment represents learning about their HIV status as in \cite{GT}. In such cases, it may be more realistic to assume that agents have private information even in the second stage. Unlike $\bar D_i$, the equilibrium neighborhood score $\pi_i^*$ is always observable to agents as it is a function of public information $S$. Thus it is plausible that agents base their decisions on the equilibrium quantity $\pi_i^*$ which signals a priori prevalence of treatment adoption in the neighborhood.

%--------------------------------------------------------------------------------------------------------------------------------------%

\paragraph{Random Coefficients Model of Potential Responses}
We put more structure on $Y_i(d,\pi)$ by using random coefficients model where we allow for a correlation between individual treatment status and random coefficients. Therefore our model can be seen as a correlated random coefficient model as in \cite{MT16} and \cite{wooldridge}.

%--------------------------------------------------------------------------------------------------------------------------------------%
% assumption이요...
\begin{assumption}[random coefficient model] \label{rc}\ \\
(i) For any $i\in \mathcal{N}_n$, $d\in\{0,1\}$ and $\pi\in[0,1]$, we have
\begin{eqnarray*}
Y_i(1,\pi) = \alpha_{1i}+\beta_{1i}\pi,\quad Y_i(0,\pi) = \alpha_{0i} + \beta_{0i}\pi
\end{eqnarray*}
where $(\alpha_{1i},\beta_{1i})$ and $(\alpha_{0i},\beta_{0i})$ are unit-specific coefficients.\\
(ii) For $S=(G,X,Z)$, unit-specific coefficients satisfy the following restrictions:
\begin{eqnarray*}
\E[\alpha_{1i}|S]=\E[\alpha_{1i}|X_i]=X_i'\alpha_1,\quad\&\quad  \E[\beta_{1i}|S]=\E[\beta_{1i}|X_i]=X_i'\beta_1
\end{eqnarray*}
and similarly,
\begin{eqnarray*}
\E[\alpha_{0i}|S]=\E[\alpha_{0i}|X_i]=X_i'\alpha_0,\quad\&\quad  \E[\beta_{0i}|S]=\E[\beta_{0i}|X_i]=X_i'\beta_0.
\end{eqnarray*}
\end{assumption}
%--------------------------------------------------------------------------------------------------------------------------------------%
Recall that $Y_i(1,\pi)$ represent $i$’s response when $i$ is given the treatment and $i$’s neighborhood score had been exogenously set to $\pi$. Under the Assumption \ref{rc} (i), such response is assumed to be linear in $\pi$ with the intercept $\alpha_{1i}$ and the slope $\beta_{1i}$ that are allowed to be different across agents. Similarly, $Y_i(0,\pi)$ is assumed to be linear in $\pi$ with the intercept $\alpha_{0i}$ and the slope $\beta_{0i}$. Note that unit-specific coefficients under the treatment, $(\alpha_{1i},\beta_{1i})$, are allowed to be different from those without the treatment, $(\alpha_{0i},\beta_{0i})$ for generality.

The assumption that $\pi$ affects the potential outcomes $Y_i(1,\pi)$ and $Y_i(0,\pi)$ in a linear way is only for convenience. It is straightforward to extend our model to include higher-order terms such as $\pi^2$, e.g., $Y_i(d,\pi)=\alpha_{d,i}+\beta_{d,i}\pi+\gamma_{d,i}\pi^2$ for $d\in \{0,1\}$. 

Unit-specific coefficients are unobservable random variables that are potentially dependent on unit's observed covariates. By Assumption \ref{rc} (ii), we assume that the observed parts of the coefficients depend on the public state variable $S=(G,X,Z)$ only through $X_i$. Importantly, this assumption implies that $Z$ is irrelevant for the random coefficients. This rules out the case that the treatment assignment vector $Z=(Z_i,Z_{-i})$ directly affects $Y_i$. This is the standard exclusion restriction of instruments. Therefore under this assumption,  $Z$ is given a status of an instrumental variable. 

The assumption that $G$ is redundant is only for convenience as we can always include network statistics such as the number of direct peers in $X_i$. Finally, that the conditional expectation is linear in $X_i$ is also for convenience as we can always allow $X_i$ to include nonlinear functions of underlying covariates.

Under Assumption \ref{rc} (ii), we can decompose the unit-specific coefficients into its mean part given $X_i$, and its deviation from mean as follows: 
\begin{eqnarray*}
\alpha_{1i}=X_i'\alpha_1+u_{1i}, &&\E[u_{1i}|S]=0,\\
\beta_{1i}=X_i'\beta_1+e_{1i},&& \E[e_{1i}|S]=0.
\end{eqnarray*}
Analogously for $D_i=0$ as well:
\begin{eqnarray*}
\alpha_{0i}=X_i'\alpha_0+u_{0i}, &&\E[u_{0i}|S]=0,\\
\beta_{0i}=X_i'\beta_0+e_{0i},&& \E[e_{0i}|S]=0.
\end{eqnarray*}
Therefore the potential outcomes can be written as
\begin{eqnarray*}
Y_i(1,\pi)=X_i'\alpha_1+u_{1i}+\pi\big(X_i'\beta_1+e_{1i} \big),\quad \E[u_{1i}|S]=\E[e_{1i}|S]=0,\\
Y_i(0,\pi)=X_i'\alpha_0+u_{0i}+\pi\big(X_i'\beta_0+e_{0i} \big),\quad \E[u_{0i}|S]=\E[e_{0i}|S]=0,
\end{eqnarray*}
%--------------------------------------------------------------------------------------------------------------------------------------%

while the observed outcome is given as follows:
\begin{eqnarray*}
Y_i=Y_i(D_i,\pi_i^*)=
\begin{cases}
X_i'\alpha_1+u_{1i}+\pi_i^*\big(X_i'\beta_1+e_{1i} \big)&\textrm{if $D_i=1$}\\
X_i'\alpha_0+u_{0i}+\pi_i^*\big(X_i'\beta_0+e_{0i} \big)&\textrm{if $D_i=0$}
\end{cases}
\end{eqnarray*}
Our model contains the four-dimensional error term: $\eta_i=(u_{1i},e_{1i},u_{0i},e_{0i})$. By construction, $\eta_i$ are uncorrelated with $S$, i.e., $\E[\eta_i|S]=0$. By having $\eta_i$, random coefficients are allowed to be heterogeneous even after controlling for relevant observed characteristics $X_i$. The importance of allowing for such unobserved heterogeneity has been emphasized in the modern program evaluation literature (See, e.g., \cite{heckmanm}, \cite{essential} and \cite{imbens}).

 %=============================================================================%
\subsection{Parameters of Interest} \label{sec:param}
In this section, we formally define our parameters of interest, the class of average casual effects. For this purpose, let us first study average potential outcomes functions. 
\paragraph{Average potential outcomes}
Under our specifications, \emph{average potential outcomes} for agents with $X_i=x$ are computed as follows: for $\pi\in[0,1]$,
\begin{eqnarray*}
\E[Y_i(1,\pi)|X_i=x]=x'\alpha_1+ (x'\beta_1)\pi,\quad \E[Y_i(0,\pi)|X_i=x]=x'\alpha_0+ (x'\beta_0)\pi.
\end{eqnarray*}
Integrating them over identically distributed $X_i$ gives the unconditional average potential outcomes. Letting $\mu_X=\E[X_i]$,
\begin{eqnarray}
\E[Y_i(1,\pi)]&=&\mu_X'\alpha_1+(\mu_X'\beta_1)\pi\\
&=&\alpha_{1m}+\beta_{1m}\pi,\label{eq-Y}\\
\E[Y_i(0,\pi)]&=&\mu_X'\alpha_0+(\mu_X'\beta_0)\pi\\
&=&\alpha_{0m}+\beta_{0m}\pi\label{eq-Y2}
\end{eqnarray}
where $(\alpha_{1m},\beta_{1m},\alpha_{0m},\beta_{0m})=(\mu_X'\alpha_1,\mu_X'\beta_1,\mu_X'\alpha_0,\mu_X'\beta_0)$. Since $\mu_X$ is identifiable from the data, identification of $(\alpha_{1m},\beta_{1m},\alpha_{0m},\beta_{0m})$ requires one to identify $(\alpha_1,\beta_1,\alpha_0,\beta_0)$. 

 $(\alpha_{1m},\alpha_{0m})$ represent the baseline mean potential outcomes when we set $\pi=0$, i.e., $(\alpha_{1m},\alpha_{0m})=(\E[Y_i(1,0)],\E[Y_i(0,0)])$. Effect of $\pi$ is captured by  $(\beta_{1m},\beta_{0m})$. 

On the other hand, $(\alpha_1,\beta_1,\alpha_0,\beta_0)$ measures the heterogeneous effect of $X_i$ on the mean potential outcomes. To see this, notice that the following equations hold:
\begin{eqnarray*}
\E[Y_i(1,\pi)|X_i=x]&=&x'\alpha_1+\pi x'\beta_1\\
&=&\E[Y_i(1,\pi)]+(x-\mu_X)'\alpha_1+\pi(x-\mu_X)'\beta_1,\\
\E[Y_i(1,\pi)|X_i=x]&=&x'\alpha_0+\pi x'\beta_0\\
&=&\E[Y_i(0,\pi)]+(x-\mu_X)'\alpha_0+\pi(x-\mu_X)'\beta_0.
\end{eqnarray*}
Therefore for $d\in\{0,1\}$, $(\alpha_d,\beta_d)$, without constant coefficients parts, explains the difference between $\E[Y_i(d,\pi)|X_i=x]$ and $\E[Y_i(d,\pi)]$.

%=============================================================================%

\paragraph{Average causal effects} Given the average response functions, we now define average causal effects, which are our parameters of interest. Let us define the \emph{average direct effect (ADE)} of own treatment under $\pi$ as follows:

$$ADE(\pi)=\E[Y_i(1,\pi)-Y_i(0,\pi)].$$
$ADE(\pi)$ measures the average change in outcomes under the regime in which $i$ is required to choose $D_i=1$, compared to the regime in which $i$ is forbidden to choose $D_i=1$ while $i$’s neighborhood score is fixed to $\pi$.  Under our random coefficients specification, $ADE(\pi)$ can be written as
 \begin{eqnarray*}
ADE(\pi)=\alpha_{1m}-\alpha_{0m}+(\beta_{1m}-\beta_{0m})\pi.
\end{eqnarray*}

%--------------------------------------------------------------------------------------------------------------------------------------%
% ASE 요...

Similarly, we define \emph{average spillover effect (ASE)} from changing the neighborhood score from $\pi$ to $\tilde\pi$ for each $d\in\{0,1\}$ as follows:
\begin{eqnarray*}
ASE(\pi,\tilde\pi,d) = \E[Y_i(d,\tilde\pi)-Y_i(d,\pi)]=(\tilde\pi-\pi)\beta_{dm},
\end{eqnarray*}
which measures the effect of changing the neighborhood score from $\pi$ to $\tilde\pi$ while fixing agent's treatment status at $D_i=d$.
Whether $\beta_{0m}=0$ or $\beta_{1m}=0$ is of interest as it indicates whether there are treatment spillovers at the outcome level.

%=============================================================================%

%=============================================================================%

\subsection{Source of Endogeneity}
In sum, our model of treatment choices and outcomes can be written as the following semi-triangular system:
\begin{eqnarray}
&&Y_i=Y_i(D_i,\pi_i^*)=
\begin{cases}
X_i'\alpha_1+u_{1i}+\big(X_i'\beta_1+e_{1i} \big)\pi_i^*&\textrm{if $D_i=1$}\\
X_i'\alpha_0+u_{0i}+\big(X_i'\beta_0+e_{0i} \big)\pi_i^*&\textrm{if $D_i=0$} \label{Ycase}
\end{cases}\\
&&D_i=\1\{v_i\le X_i'\theta_1+\theta_2 Z_i+\theta_3\pi_i^*\}\label{dmodel}\\
&&\textrm{  s.t. } \sigma_i^{*}=\Phi\Big(X_i'\theta_1+\theta_2 Z_i+\theta_3\pi_i^{*}\Big),\quad \forall i\in\mathcal{N}_n.
\end{eqnarray}
Using the formula $Y_i=D_iY_i(1,\pi_i^*)+(1-D_i)Y_i(0,\pi_i^*)=Y_i(0,\pi_i^*)+D_i(Y_i(1,\pi_i^*)-Y_i(0,\pi_i^*))$, \ref{Ycase} can be written as follows:

\begin{eqnarray}
Y_i=X_i'\alpha_0+\pi_i^*X_i'\beta_0+ D_iX_i'(\alpha_1-\alpha_0)+D_i\pi_i^*X_i'(\beta_1-\beta_0)+\epsilon_i\label{YY}
\end{eqnarray}
where
\begin{eqnarray}
\epsilon_i=u_{0i}+\pi_i^*  e_{0i}+D_i\big(u_{1i}-u_{0i}+\pi_i^*(e_{1i}-e_{0i})\big).\label{EEE}
\end{eqnarray}

Equation \ref{YY} gives the conventional linear regression model. Naturally, one may consider estimating $(\alpha_1,\alpha_0,\beta_1,\beta_0)$ by the least squares regression of $Y_i$ on $(X_i,\pi_i^*X_i,D_iX_i,D_i\pi_i^*X_i)$. Resulting OLS estimator is consistent only when $\epsilon_i$ is uncorrelated with the regressors, i.e.,  $\E[\epsilon_i|D_i,X_i,\pi_i^*]=0$ which requires that the following two conditions hold:
\begin{eqnarray*}
&&\E[u_{0i}+\pi_i^*e_{0i}|D_i=0,X_i,\pi_i^*]=^{(a)}\E[u_{0i}+\pi_i^*e_{0i}|X_i,\pi_i^*]=^{(b)}0,\\
&&\E[u_{1i}+\pi_i^*e_{1i}|D_i=1,X_i,\pi_i^*]=^{(a)'}\E[u_{1i}+\pi_i^*e_{1i}|X_i,\pi_i^*]=^{(b)'}0.
\end{eqnarray*}
Since $\eta_i=(u_{1i},u_{0i},e_{1i},e_{0i})$ are uncorrelated with $S=(G,X,Z)$ by construction, $(b)$ and $(b)'$ are automatically satisfied. Therefore, we only need to show that $(a)$ and $(a)'$ are satisfied. This is true only when $D_i$ is uncorrelated with $\eta_i$ conditional on $(X_i,\pi_i^*)$. This is the familiar selection-on-observables assumption. Such assumption is unlikely to hold if the treatment group and control group are systematically different in their unobserved factors $\eta_i$ even after controlling for all relevant observables. Indeed, the very fact that agents with the same observed characteristics $(X_i,\pi_i^*)$ have made different treatment choices suggests that they differ in their unobserved factors. Thus, the source of endogeneity comes from the correlation between $v_i$ and $\eta_i$ even after conditional on $S$.

More specifically, note that the selection-on-observables assumption requires that the following two conditions hold:
\begin{eqnarray}\label{sel1}
Corr(Y_i(0,\pi_i^*),D_i|X_i,\pi_i^*)=0
\end{eqnarray}
and
\begin{eqnarray}\label{sel2}
Corr(Y_i(1,\pi_i^*)-Y_i(0,\pi_i^*),D_i|X_i,\pi_i^*)=0.
\end{eqnarray}

Condition \ref{sel1} requires that the idiosyncratic part of $Y_i(0,\pi_i^*)$ is uncorrelated with $D_i$,
i.e., in the absence of the treatment, there should be no difference in the mean potential outcomes across treatment group and control group once we account for relevant observables $(X_i,\pi_i^*)$. However, agents who take up the treatment may have unusual values of $Y_{i}(0,\pi)$ even after controlling for $(X_i,\pi_i^*)$. If individuals who take up the treatment tend to have higher values of $Y_i(0,\pi)$ in terms of unobservables, then the naive least squares regression would suffer from an upward bias since $cov(D_i,\epsilon_i|S)>0$. This is the case of classic selection problem. 

The requirement \ref{sel2} is also troublesome as the condition implies that the unobserved gain from the treatment given $\pi_i^*$ should not vary across treatment group and control group. This is not satisfied if the treatment choice is correlated with unobserved gains from the treatment. It is plausible that agents have some knowledge of likely idiosyncratic gains from the treatment at the time they choose their treatment status. If agent's treatment choice is partially based on such knowledge, then \ref{sel2} would not be satisfied. This type of sorting on the unobserved gain, termed ``\emph{essential heterogeneity}" by \cite{essential},  has been emphasized in the modern program literature.

In conclusion, whenever selection problem or essential heterogeneity exists, the naive OLS regression delivers inconsistent estimates of structural parameters $(\alpha_1,\alpha_0,\beta_1,\beta_0)$. 

%=============================================================================%

\section{Identification}\label{sec:idd}
In the previous section, we showed that the OLS regression of \ref{YY} suffers from bias when $v_i$ is correlated with $\eta_i=(u_{1i},u_{0i},e_{1i},e_{0i})$ even when we control for $S$. In this section, we first show that the IV methods do not identify the casual parameters of interest in the presence of general heterogeneity. We then propose the alternative method known as control function approach.

\subsection{The Problem of Conventional IV Methods}
Endogeneity is often addressed by IV methods such as two-stage least squares (2SLS). In our setup, $Z_i$ is a valid IV for $D_i$ since (i) $D_i$ is correlated with $Z_i$, and (ii) $Z_i$ is exogenous and is excluded from the outcome equation. In fact, in the presence of spillovers in the first stage, not only $Z_i$ but also $n$-dimensional vector $Z=(Z_i,Z_{-i})$ is a valid instrument for $D_i$ since in that case, $D_i$ is a function of entire assignment vector $Z$.\footnote{Recall that when there exist spillovers in the first stage choice model, not only $i$'s direct neighbor's $Z$ but indirect neighbors' $Z$ also affect $D_i$. Therefore $Z_j$ for $j$ that are eventually connected to $i$ is also relevant for $D_i$. However as the network distance between $i$ and $j$ becomes greater, the dependence between $Z_j$ and $D_i$ decays exponentially when $\lambda<1$. (See \cite{xu18} and \cite{L20}). Therefore, using $Z_j$ that is too far from $i$ as an IV may incur weak IV problem.}. Therefore, we may run an IV regression to \ref{YY} where we instrument  $D_i$ by $Z_i$ or by $Z=(Z_i,Z_{-i})$, depending on whether spillovers exist in the first stage. 

We argue that such strategy does not identify $(\alpha_0,\beta_0,\alpha_1,\beta_1)$ in our setup.  Suppose we instrument $D_i$ by $Z_i$. 
The resulting IV estimator is consistent only when the $\E[\epsilon_i|Z_i,X_i,\pi_i^*]=0$ where $\epsilon_i=u_{0i}+\pi_i^*  e_{0i}+D_i\big(u_{1i}-u_{0i}+\pi_i^*(e_{1i}-e_{0i})\big)$ as in \ref{EEE}. Note that,
\begin{eqnarray*}
&&\E[\epsilon_i|Z_i,X_i,\pi_i^*]\\&=&\E[u_{0i}+\pi_i^*  e_{0i}+D_i\big(u_{1i}-u_{0i}+\pi_i^*(e_{1i}-e_{0i})\big)|Z_i,X_i,\pi_i^*]\\
&=&\underbrace{\E[u_{0i}+\pi_i^*  e_{0i}|Z_i,X_i,\pi_i^*]}_{\textbf{A}}+\underbrace{\E[ u_{1i}-u_{0i}+\pi_i^*(e_{1i}-e_{0i})|D_i=1,Z_i,X_i,\pi_i^*]}_{\textbf{B}}\underbrace{\Pr(D_i=1|Z_i,X_i,\pi_i^*)}_{\textbf{C}}.
\end{eqnarray*}
$\textbf{A}=0$ since $\eta_i=(u_{1i},u_{0i},e_{1i},e_{0i})$ is uncorrelated with $S$, and thereby with $(Z,X_i,\pi_i^*)$. $\textbf{C}$ cannot be zero except for trivial cases. Therefore $\E[\epsilon_i|Z,X_i,\pi_i^*]=0$ only when $\textbf{B}=0$. This is satisfied when $\E[u_{1i}-u_{0i}+\pi_i^*(e_{1i}-e_{0i})|D_i=1,Z_i,X_i,\pi_i^*]=\E[u_{1i}-u_{0i}+\pi_i^*(e_{1i}-e_{0i})|Z_i,X_i,\pi_i^*]$ as $\E[\eta_i|S]=0$ implies that the last term is zero. Note that $u_{1i}-u_{0i}+\pi_i^*(e_{1i}-e_{0i})$ can be interpreted as an idiosyncratic part of $Y_i(1,\pi_i^*)-Y_i(0,\pi_i^*)$. Therefore we need to assume that $D_i$ is uncorrelated with the idiosyncratic gain from taking the treatment once we condition on $(Z_i,X_i,\pi_i^*)$. Such requirement is unrealistic when agents have some knowledge on their idiosyncratic gains and base their treatment  decision on such knowledge, i.e., when there is sorting on unobserved gains.

Whether the $u_{1i}-u_{0i}+\pi_i^*(e_{1i}-e_{0i})$ is correlated with $D_i$ is an empirical matter and should not be settled a priori. IV methods rule out the possibility of such correlation and are subject to failure when the correlation exists. This point has also been pointed out in the traditional  treatment effect literature which rules out spillover effects. (See \cite{HR}). For instance, it is now well established in the literature that IV/2SLS does not recover the average causal parameters such as ATE under the heterogeneous responses model such as random coefficients models (See \cite{IA}).

%--------------------------------------------------------------------------------------------------------------------------------------%

\subsection{Control Function Approach}
%--------------------------------------------------------------------------------------------------------------------------------------%
We now propose the alternative strategy known as the control function approach. Control function approach addresses the endogeneity problem by explicitly formulating the dependence between outcomes and treatments. To apply this method, we first write the observed conditional means $\E[Y_i|D_i=1,S]$ and $\E[Y_i|D_i=0,S]$ as follows:
\begin{eqnarray*}
&&\E[Y_i|D_i=1,S]\\
&=&\E[Y_i|D_i=1,\sigma_i^*(S,\theta),\pi_i^*(S,\theta),S]\\
&=&\E[Y_i(1,\pi_i^*(S,\theta))|v_i\le\Phi^{-1}(\sigma_i^*(S,\theta)),\sigma_i^*(S,\theta),\pi_i^*(S,\theta),S]\\
&=&X_i'\alpha_1+\E[u_{1i}|v_i\le \Phi^{-1}(\sigma^*_i(S,\theta)),S]+\pi_i^*(S,\theta)\Big\{X_i'\beta_1+\E[e_{1i}|v_i\le \Phi^{-1}(\sigma^*_i(S,\theta)),S]\Big\}
\end{eqnarray*}
since $D_i=1$ $\Longleftrightarrow$ $\Phi(v_i)\le \sigma_i^*$ (See \ref{Dp3}). Similarly, the observed conditional mean for the control group is,
\begin{eqnarray*}
&&\E[Y_i|D_i=0,S]\\
&=&X_i'\alpha_0+\E[u_{0i}|v_i> \Phi^{-1}(\sigma^*_i(S,\theta)),S]+\pi_i^*(S,\theta)\Big\{X_i'\beta_0+\E[e_{0i}|v_i> \Phi^{-1}(\sigma^*_i(S,\theta)),S]\Big\}.
\end{eqnarray*}

The terms $\E[u_{1i}|v_i\le \Phi^{-1}( \sigma^*_i(S,\theta)),S], \E[e_{1i}|v_i\le \Phi^{-1}(\sigma^*_i(S,\theta)),S]$ and $\E[u_{0i}|v_i>\Phi^{-1}( \sigma^*_i(S,\theta)),S],\E[e_{0i}|v_i>\Phi^{-1}( \sigma^*_i(S,\theta)),S]$ are ``control functions" which account for the endogeneity of $D_i$. Assumption \ref{eta} below restricts the form of these control functions.
 
\begin{assumption}\label{eta}  For all $i\in\mathcal{N}_n$, $\eta_i=(u_{1i},u_{0i},e_{1i},e_{0i})$ satisfies the following conditions.
\begin{enumerate}
\item[(i)] $\eta_i$ is i.i.d. and is independent of $S$.
\item[(ii)] $\E[\eta_i|v_i]$ is a linear function of $v_i$. 
\end{enumerate}
Under these two conditions, we write
\begin{eqnarray*}
\E[u_{1i}|v_i,S]=\E[u_{1i}|v_i]=\rho_{u_1}v_i,&& \E[e_{1i}|v_i,S]=\E[e_{1i}|v_i]=\rho_{e_1}v_i,\\
\E[u_{0i}|v_i,S]=\E[u_{0i}|v_i]=\rho_{u_0}v_i,&& \E[e_{0i}|v_i,S]=\E[e_{0i}|v_i]=\rho_{e_0}v_i
\end{eqnarray*}
where $\rho=(\rho_{u_1},\rho_{e_1},\rho_{u_0},\rho_{e_0})$ captures the covariances between each component of $\eta_i$ and $v_i$.
\end{assumption}

Assumption \ref{eta} (i) is often referred to as ``separability" assumption and has been utilized in literature as in \cite{carneiro} and \cite{brinch}.  Under this assumption, the control functions depend only on the individual propensity score $\sigma_i^*(S,\theta)$, e.g., $\E[u_{1i}|v_i\le\Phi^{-1}(\sigma_i^*(S,\theta)),S]=\E[u_{1i}|v_i\le\Phi^{-1}(\sigma_i^*(S,\theta))]$ so that the control functions are separated from $S$. As a result, $\E[Y_i|D_i=1,S]$ and $\E[Y_i|D_i=0,S]$ depend on $S$ only though $(X_i,\pi_i^*,\sigma_i^*)$. This step is necessary since it is not possible to control for $S=(G,X,Z)$ itself as our data consist of one large network. 

Assumption \ref{eta} (ii) further allows us to write $\E[u_{1i}|v_i\le\Phi^{-1}(\sigma_i^*)]$, for instance, as $\rho_{u_1}\E[v_i|v_i\le\Phi^{-1}(\sigma_i^*)]$. Combined with the normality assumption on $v_i$, we effectively assume that $(\eta_i,v_i)$ are jointly normal. However, it can easily accommodate alternative distributional assumptions on $v_i$ other than normality.

Under the joint normality assumption, control functions take a form of inverse mills ratio. 
Define $\lambda_1(\cdot)$ and $\lambda_0(\cdot)$ as follows: For $\sigma\in (0,1)$, 
 $$\lambda_1(\sigma)=-\frac{\phi(\Phi^{-1}(\sigma))}{\sigma},\quad \lambda_0(\sigma)=\frac{\phi(\Phi^{-1}(\sigma))}{1-\sigma}. $$ 
It follows that
\begin{eqnarray*}
&&\E[Y_i|D_i=1,S]=X_i'\alpha_1+\rho_{u_1}\lambda_1(\sigma_i^*)+\pi_i^*\big(X_i'\beta_1+\rho_{e_1}\lambda_{1}(\sigma_i^*) \big),\\
&&\E[Y_i|D_i=0,S]=X_i'\alpha_0+\rho_{u_0}\lambda_0(\sigma_i^*)+\pi_i^*\big(X_i'\beta_0+\rho_{e_0}\lambda_{0}(\sigma_i^*) \big).
\end{eqnarray*}
Let $\lambda_i=D_i\lambda_{1i}+(1-D_i)\lambda_{0i}$. We see that $(\alpha_1,\beta_1,\rho_{u_1},\rho_{e_1})$ is identified by regressing $Y_i$ on $(X_i',\lambda_i,\pi_i^*X_i',\pi_i^*\lambda_i)'$ using the subsample of $D_i=1$. Similarly, we can identify $(\alpha_0,\beta_0,\rho_{u_0},\rho_{e_0})$ by regressing $Y_i$ on $X_i,\lambda_i$ and their interactions with $\pi_i^*$ using the subsample of $D_i=0$. The inclusion of $\lambda_i$ accounts for the correlation between $\eta_i$ and $v_i$ so that we can test for the endogeneity of $D_i$ by checking whether correlations are collectively zero or not.

Our model achieves a point identification by exploiting a functional form assumption between $\eta_i$ and $v_i$. We can relax the linearity assumption and have more flexible parametric functional form by adding higher-order terms. For instance, we may specify $\E[u_{1i}|v_i]$ as the quadratic function of $v_i$ as follows:
$$\E[u_{1i}|v_i]=\rho_{u_1}v_i+\tilde\rho_{u_1}v_i^2. $$
Then it can be shown that
\begin{eqnarray*}
\E[u_{1i}|v_i\le \Phi^{-1}(\sigma_i^*)]&=& -\rho_{u_1}\frac{\phi(\Phi^{-1}(\sigma_i^*))}{\sigma_i^*} +\tilde\rho_{u_1}\Big[
\Phi^{-1}(\sigma_i^*)\frac{\phi(\Phi^{-1}(\sigma_i^*))}{\sigma_i^*} 
+ \Big\{\frac{\phi(\Phi^{-1}(\sigma_i^*))}{\sigma_i^*}\Big\}^2\Big].
\end{eqnarray*}
This also offers a way to test for linearity assumption in a spirit of \cite{lee}.

%--------------------------------------------------------------------------------------------------------------------------------------%
%--------------------------------------------------------------------------------------------------------------------------------------%
%=============================================================================%
%s_estimation
%=============================================================================%

\section{Estimation}\label{sec:estimation}
We propose a two-stage estimation procedure. In the first-stage, we estimate the treatment choice games using a nested fixed point maximum likelihood (NFXP-ML) method. In the second-stage, using first-stage estimates, we estimate regression models of treatment outcomes with generated regressors.

\subsection{First-Stage Estimation}
Recall that the treatment choice models boil down to equation \ref{dmodel} subject to the fixed-point requirement \ref{fpoint}. Our sample log-likelihood function are defined as follows:
\begin{eqnarray}
\widehat{\mathcal{L}}_n(\theta)&=&\frac{1}{n}\sum_{i=1}^n \Big\{D_i\ln \sigma_i^*(S,\theta)+(1-D_i)\ln (1-\sigma_i^*(S,\theta))\Big\} \label{loglike}
\end{eqnarray}
Our estimator $\hat\theta=(\hat\theta_1,\hat\theta_2,\hat\theta_3)$ is defined as the maximizer of $\widehat{\mathcal{L}}_n(\theta)$ subject to the constraint that $\{\sigma_i^*(S,\hat\theta)\}$ satisfies the fixed-point requirement. Formally,

\begin{eqnarray}
\hat\theta=\arg\max_{\theta\in\Theta} \widehat{\mathcal{L}}_n(\theta)
\end{eqnarray}
subject to
\begin{eqnarray}
\sigma_i^{*}(S,\hat\theta)=\Phi\Big(X_i'\hat\theta_1+\hat\theta_2 Z_i+\hat\theta_3\frac{1}{|\mathcal{N}_i|}\sum_{j\in\mathcal{N}_i}\sigma_j^{*}(S,\hat\theta)\Big),\quad \forall i\in \mathcal{N}_n \label{fpoint}
\end{eqnarray}
For computation, we use the nested fixed point (NFXP) algorithm. Specifically, starting with an arbitrary initial guess for $\hat\theta$, we find the fixed point of \ref{fpoint} via contraction iterations (it can be shown that \ref{fpoint} is a contraction mapping when $\lambda<1$).
We then compute the log-likelihood function \ref{loglike} using the obtained conditional choice probabilities. Update $\hat\theta$ to $\hat\theta'$ according to, say, Newton's method. Iterate the procedure until a sequence of estimates converges. Our NFXP-ML estimator is taken as its limit.

%--------------------------------------------------------------------------------------------------------------------------------------%
\subsection{Second-Stage Estimation}
Let us define the set of regressors as
$$W_i=[X_i', \lambda_i,\pi_i^*(S,\theta)X_i',\pi_i^*(S,\theta)\lambda_i]' $$
where $\lambda_i=D_i\lambda_{1i}+(1-D_i)\lambda_{0i}$ with $\lambda_{1i}=\lambda_1(\sigma_i^*(S,\theta))$ and $\lambda_{0i}=\lambda_0(\sigma_i^*(S,\theta))$. 

Our estimators are based on the following moment conditions
\begin{eqnarray*}
\E[Y_i|D_i=1,S]=W_{i}'\gamma_1,\quad \E[Y_i|D_i=0,S]=W_i'\gamma_0
\end{eqnarray*}
where $\gamma_1=(\alpha_1,\rho_{u_1},\beta_1,\rho_{e_1})'$ and $\gamma_0=(\alpha_0,\rho_{u_0},\beta_0,\rho_{e_0})'$.

This suggests that $\gamma_1$ and $\gamma_0$ can be estimated by regressing $Y_i$ on $W_i$, separately to the subsample with $D_i=1$ and $D_i=0$, respectively. However, since $\lambda_i$ and $\pi_i^*$ are functions of unknown first-stage parameters $\theta$, we need to replace $\theta$ with $\hat\theta$.  Define $\hat\lambda_{1i}=\lambda_{1}(\sigma_i^*(S,\hat\theta))$ and $\hat\lambda_{0i}=\lambda_0(\sigma_i^*(S,\hat\theta))$. Let $\hat\lambda_{i}=D_i\hat\lambda_{1i}+(1-D_i)\hat\lambda_{0i}$. Similarly, we replace the unknown quantity $\pi_i^*(S,\theta)\equiv\frac{1}{|\mathcal{N}_i|}\sum_{j\in \mathcal{N}_i}\sigma_j^*(S,\theta)$ with $\hat\pi_i^*=\pi_i^*(S,\hat\theta)=\frac{1}{|\mathcal{N}_i|}\sum_{j\in\mathcal{N}_i}\sigma_j^*(S,\hat\theta)$. Thus, our generated regressor $\hat W_i$ for $W_i$ is
$$ \hat W_i=[X_i', \hat\lambda_i,\hat\pi_iX_i',\hat\pi_i\hat\lambda_i]'.$$

Estimator for $\gamma_1$ is then defined as 
\begin{eqnarray*}
\hat\gamma_1 &=& \arg\min_{\gamma_1} \frac{1}{n}\sum_{i=1}^n D_i\big(Y_i-\hat W_i'\gamma_1\big)^2 \\
&=&\Big\{\sum_{i=1}^n D_i\hat W_i\hat W_i'\Big\}^{-1}\sum_{i=1}^nD_i\hat W_iY_i.
\end{eqnarray*}
Similarly, estimator for $\gamma_0$ is
\begin{eqnarray*}
\hat\gamma_0 &=& \arg\min_{\gamma_0} \frac{1}{n}\sum_{i=1}^n (1-D_i)\big(Y_i-\hat W_i'\gamma_0\big)^2 \\
&=&\Big\{\sum_{i=1}^n (1-D_i)\hat W_i\hat W_i'\Big\}^{-1}\sum_{i=1}^n(1-D_i)\hat W_iY_i.
\end{eqnarray*}

%--------------------------------------------------------------------------------------------------------------------------------------%
\subsection{Inference}

For the asymptotic analysis, we consider large-network asymptotics in which a number of individuals connected in a single network goes to infinity. Moreover, for each $n$, we treat $S=(G,X,Z)$ as fixed. This is justified since $S$ is an ancillary statistics, i.e., $S$ does not contain any information on the parameters of interest.

%--------------------------------------------------------------------------------------------------------------------------------------%

%--------------------------------------------------------------------------------------------------------------------------------------%
\subsubsection{Inference for the first-stage game}
We first establish $\sqrt{n}$-consistency and asymptotic normality of the first-stage estimator $\hat\theta$. The true parameter is denoted by $\theta^0$. Therefore our data $\{D_i\}_{i=1}^n$ is assumed to be generated from 
$$D_i=\1\{v_i\le X_i'\theta_1^0+\theta_2^0 Z_i+\theta_3^0\pi_i^*(S,\theta^0)\} $$
subject to $\sigma_i^*(S,\theta^0)=\Phi\big(X_i'\theta_1^0+\theta_2^0 Z_i+\theta_3^0\pi_i^*(S,\theta^0)\big)$ for all $i\in \mathcal{N}_n$. 

%--------------------------------------------------------------------------------------------------------------------------------------%

\begin{theorem}[consistency of $\hat\theta$]\label{thm:A} Under the following assumptions, $\hat\theta-\theta^0\xrightarrow{p}   0$.
\begin{thmlist}
\item The true parameter $\theta^0=(\theta_1^0,\theta_2^0,\theta_3^0)$ lies in a compact set $\Theta\subseteq \mathbb{R}^{dim(\theta)}$ and $|\theta_3^0|<\sqrt{2\pi}$. The support of $X_i$ is a bounded subset of $\mathbb{R}^{k}$. \label{thm:A2}
\item Let $R_i=(X_i',Z_i,\pi_i^*(S,\theta^0))'$. For large enough $n$, $\sum_{i=1}^nR_iR_i’$ is invertible, i.e., 
$$\liminf_{n\to\infty} \det(\sum_{i=1}^n R_iR_i’ )>0.$$  \label{thm:A3}
\end{thmlist}
See Appendix \ref{thm_consistency_FS} for the proof.
\end{theorem}
%--------------------------------------------------------------------------------------------------------------------------------------%
Assumption \ref{thm:A2} ensures that there is unique equilibrium at the true parameter (See Theorem \ref{ue}) and that each equilibrium probability $\sigma_i^*(S,\theta)\in(0,1)$ for all $i$. 
 Assumption \ref{thm:A3} is the rank condition for identification which requires that for all large enough $n$. the moment matrix of regressors has full rank.

%--------------------------------------------------------------------------------------------------------------------------------------%
We now establish asymptotic normality of $\hat\theta$. Let us define the information matrix as follows:
$$\mathcal{I}_n(\theta)=\E\Big[\frac{1}{n}\sum_{i=1}^n\nabla_{\theta} l_i(\theta)\nabla_{\theta} l_i(\theta)'\Big|S\Big] $$
where $l_i(\theta)=D_i\ln\sigma_i^*(S,\theta)+(1-D_i)\ln(1-\sigma_i^*(S,\theta)) $ is the individual log-likelihood function. Therefore $\nabla_\theta l_i(\theta)$ is given by
\begin{eqnarray}\label{score2}
\nabla_\theta l_i(\theta)= D_i\frac{\nabla_\theta \sigma_i^*(S,\theta)}{\sigma_i^*(S,\theta)}+(1-D_i)\frac{-\nabla_\theta\sigma_i^*(S,\theta)}{1-\sigma_i^*(S,\theta)}.
\end{eqnarray}
%--------------------------------------------------------------------------------------------------------------------------------------%
\begin{theorem}[asymptotic normality of $\hat\theta$] In addition to the conditions for Theorem \ref{thm:A}, assume
\begin{thmlist}
\item The true parameter $\theta^0$ lies in the interior of the compact set $\Theta\subseteq \mathbb{R}^{dim(\theta)}$. \label{thm:B1}
\item For any $n$, $\mathcal{I}_n(\theta^0)$ is nonsingluar.  \label{thm:B2}
\end{thmlist}
 Then
\begin{eqnarray}
(\mathcal{I}_n^{-1}(\theta^0))^{-1/2}\sqrt{n}(\hat\theta-\theta^0)\xrightarrow{d}   N(0,I_{dim(\theta)}) \label{FSnorm}
\end{eqnarray}
where $I_{dim(\theta)}$ is the $dim(\theta)\times dim(\theta)$ identity matrix.\\ 
See Appendix \ref{thm_norm_FS} for proof.
\end{theorem}

%--------------------------------------------------------------------------------------------------------------------------------------%

\paragraph{Variance Estimation} The asymptotic variance of $\hat\theta$ can be estimated by $\widehat{Var}(\hat\theta)=\widehat{\mathcal{I}}_n^{-1}/n$ where

$$\widehat{\mathcal{I}}_n \equiv \frac{1}{n}\sum_{i=1}^n \nabla_{\theta} l_i(\hat\theta)\nabla_{\theta} l_i(\hat\theta)'. $$
In order to compute $\nabla_\theta l_i(\hat\theta)$ using equation \ref{score2}, we need to evaluate $\nabla_\theta \sigma_i^*(S,\hat\theta)$. For this we use the numerical approximation method:  Take $\hat\theta+\epsilon$ for a small perturbation $\epsilon$ (e.g., $\epsilon=10^{-5})$, then compute the new equilibrium  $\{\sigma_i^*(S,\hat\theta+\epsilon)\}_{i=1}^n$ by solving the fixed point. $\nabla_\theta\sigma_i^*(S,\hat\theta)$ is then computed by $(\sigma_i^*(S,\hat\theta+\epsilon)-\sigma_i^*(S,\hat\theta))/\epsilon$.

%--------------------------------------------------------------------------------------------------------------------------------------SS inference% %--------------------------------------------------------------------------------------------------------------------------------------%
\subsubsection{Inference for second-stage regression} Next, we establish $\sqrt{n}$-consistency and asymptotic normality of the second-stage estimators $(\hat\gamma_1,\hat\gamma_0)$. Let us denote the true parameters by $(\gamma_1^0,\gamma_0^0)$. We assume that our model is correctly specified, i.e., $Y_i$ satisfies the following conditional moment restrictions:
$$\E[Y_i|S,D_i=1]=W_i'\gamma_1^0,\quad\E[Y_i|S,D_i=0]=W_i'\gamma_0^0. $$
We maintain the conditions for $\sqrt{n}$-consistency and asymptotic normality of the first-stage estimator $\hat\theta$.
%--------------------------------------------------------------------------------------------------------------------------------------%

\begin{theorem}[consistency of $(\hat\gamma_1,\hat\gamma_0)$]\label{con2} Under the following assumptions, $\hat\gamma^0_1-\gamma^0_1\xrightarrow{p}   0$ and $\hat\gamma^0_0-\gamma^0_0\xrightarrow{p}   0$ 
\begin{thmlist}
\item The true parameter $\gamma_1^0$ lies in a compact set $\Gamma_1\subseteq \mathbb{R}^{dim(\gamma_1)}$. Similarly, the true parameter $\gamma_0^0$ lies in a compact set $\Gamma_0\subseteq \mathbb{R}^{dim(\gamma_0)}$.
\item\label{thm:B2} Let
$$\liminf_{n\to\infty}\det\Big\{\sum_{i=1}^n\E[D_iW_iW_i'|S] \Big\}>0 $$
and
$$\liminf_{n\to\infty}\det\Big\{\sum_{i=1}^n\E[(1-D_i)W_iW_i'|S] \Big\}>0. $$
\end{thmlist}
See Appendix \ref{thm_consistency} for proof.
\end{theorem}

%--------------------------------------------------------------------------------------------------------------------------------------%
%--------------------------------------------------------------------------------------------------------------------------------------%

Next, we derive the asymptotic results for the second-step estimators. For compactness, we only report results for $\hat\gamma_1$, as $\hat\gamma_0$ case can be derived in an analogous way.

\begin{theorem}[asymptotic normality of $\hat\gamma_1$] Define
\begin{eqnarray*}
\Upsilon_n&=&\E[\frac{1}{n}\sum_{i=1}^n D_iW_iW_i'|S]\\
\Psi_n&=&\E[\frac{1}{n}\sum_{i=1}^nD_i W_iW_i'\epsilon_{1i}^2|S]\\
&+&\E\Big[\frac{1}{n}\sum_{i=1}^n D_iW_i\gamma_1^{0\prime}\nabla_{\gamma_1} W_i(\gamma_1^0)\Big|S\Big]
\E\Big[\frac{1}{n}\sum_{i=1}^n\nabla_{\theta} l_i(\theta^0)\nabla_{\theta} l_i(\theta^0)'\Big|S\Big]^{-1}
\E\Big[\frac{1}{n}\sum_{i=1}^n D_iW_i\gamma_1^{0\prime}\nabla_{\gamma_1} W_i(\gamma_1^0)\Big|S\Big]'
\end{eqnarray*}
In addition to the conditions for Theorem \ref{con2}, assume
\begin{thmlist}
\item The true parameter $\gamma_1^0$ lies in the interior of the compact set $\Gamma_1\subseteq \mathbb{R}^{dim(\gamma_1)}$. 
\item For any $n$, $\Psi_n$ and $\Upsilon_n$ are nonsingular. 
\end{thmlist}
Then we have
$$\Lambda_n^{-1/2}\sqrt{n}(\hat\gamma_1-\gamma_1^0)\xrightarrow{d}   N(0,I_{dim(\gamma_1)}) $$
where $\Lambda_n=\Upsilon_n^{-1}\Psi_n\Upsilon_n^{-1}$. See Appendix \ref{thm_norm_SS} for proof.
\end{theorem}
%--------------------------------------------------------------------------------------------------------------------------------------%
If we ignore first-stage estimation, the asymptotic variance would be 
$$\Upsilon_n^{-1}\E\Big[\frac{1}{n}\sum_{i=1}^nD_i W_iW_i'\epsilon_{1i}^2|S\Big]\Upsilon_n^{-1}$$
 which is smaller, in the positive semi-definite sense, than the correct asymptotic variance $\Upsilon_n^{-1}\Psi_n\Upsilon_n^{-1}$.

\paragraph{Variance Estimation} The asymptotic variance $\Lambda_n$ can be estimated by replacing the population means by sample counterparts. Specifically,
\begin{eqnarray*}
\hat\Upsilon_n&=&\frac{1}{n}\sum_{i=1}^n D_i\hat W_i\hat W_i'\\
\hat\Psi_n&=&\frac{1}{n}\sum_{i=1}^n D_i \hat W_i\hat W_i'\hat\epsilon_{1i}^2
+ \Big(\frac{1}{n}\sum_{i=1}^n D_iW_i\hat\gamma_1'\nabla_{\gamma_1} W_i(\hat\gamma_1)\Big)\Big(\frac{1}{n}\sum_{i=1}^n  \nabla_\theta l_i(\hat\theta)\nabla_\theta l_i(\hat\theta)  \Big)
 \Big(\frac{1}{n}\sum_{i=1}^n D_iW_i\hat\gamma_1'\nabla_{\gamma_1} W_i(\hat\gamma_1)\Big)'\\
\end{eqnarray*}
where $\hat\epsilon_{1i}=D_i(Y_i-\hat W_i'\hat\gamma_1)$.

%--------------------------------------------------------------------------------------------------------------------------------------%
\subsection{Monte Carlo Simulation}

In this section, we illustrate the finite sample properties of our estimators through simulation exercises. 

\paragraph{Exogenous Variables} For simulation purpose, we imitate the environment of \cite{dupas}.  The network $G$ is constructed from the GPS data of \cite{dupas}. Specifically, two households $i$ and $j$ are considered connected if they live within 500-meter radius. After removing isolated nodes, we have a sample size of 538. The instrumental variable $Z$ is also taken from \cite{dupas} where the binary $Z_i$ represents whether $i$ received a high level of subsidy or not. Summary statistics of $(G,Z)$ can be found in the next section. Throughout the simulation replications, $G$ and $Z$ are treated fixed. We do not consider $X$.

\paragraph{Generating Endogenous Variables}
Treatment choices are determined according to the following equation:
$$D_i=\1\{v_i\le \theta_1+\theta_2 Z_i+\theta_3\pi_i^*\} $$ 
where $v_i\sim^{iid}N(0,1)$. We set $\theta=(\theta_1,\theta_2,\theta_3)=(-2,1,1.5)$ under which the probability of $D=1$ is around 0.8. Since $|\theta_3|<2.5$, there exists a unique equilibrium by the Theorem \ref{ue}. Given our parameter values, we can compute the unique equilibrium $\{\sigma_i^*(G,Z,\theta)\}_{i=1}^n$ by calculating the fixed point to the following system:
$$\sigma_i^*(G,Z,\theta)=\Phi\big\{\theta_1+\theta_2 Z_i+\theta_3\frac{1}{|\mathcal{N}_i|}\sum_{j\in\mathcal{N}_i} \sigma_j^*(G,Z,\theta)\big\},\quad\forall i\in\mathcal{N}_n $$
$\pi_i^*$ is then computed by $\pi_i^*(G,Z,\theta)=\sum_{j\in\mathcal{N}_i}\sigma_j^*(G,Z,\theta)/|\mathcal{N}_i|$.

Outcomes are realized according to the following rule:
\begin{eqnarray*}
Y_i=
\begin{cases}
\alpha_{1i}+\beta_{1i}\pi_i^*&\textrm{if $D_i=1$}\\
\alpha_{0i}+\beta_{0i}\pi_i^*&\textrm{if $D_i=0$}.
\end{cases}
\end{eqnarray*}
We generate the random coefficients according to
\begin{eqnarray*}
\alpha_{1i}|v_i\sim^{iid}N(2+0.3 v_i,1),\quad\beta_{1i}|v_i\sim^{iid}N(1+0.4 v_i,1),\\
\alpha_{0i}|v_i\sim^{iid}N(4+0.2 v_i,1),\quad\beta_{0i}|v_i\sim^{iid}N(3+0.2 v_i,1),
\end{eqnarray*}
so that $(\E[\alpha_{1i}],\E[\beta_{1i}],\E[\alpha_{0i}],\E[\beta_{0i}])$ or $(\alpha_1,\beta_1,\alpha_0,\beta_0)$ is given as $(2,1,4,3)$. Correlations between $(\alpha_{1i},\beta_{1i},\alpha_{0i},\beta_{0i})$ and $v_i$ are given by $(\rho_{\alpha_1},\rho_{\beta_1},\rho_{\alpha_0},\rho_{\beta_0})=(0.3,0.4,0.2, 0.2)$ so that $D_i$ is endogenous with respect to all coefficients. 
% \usepackage{graphicx}
	⁃	\begin{table}[]
\centering
\begin{tabular}{ll|ccc}
\hline
   &   coeff.     & bias   & se    & cov.prob. \\ \hline
FS & $\theta_1$ & 0.007  & 0.276 & 0.948     \\
   & $\theta_2$ & -0.034 & 0.181 & 0.937     \\
   & $\theta_3$ & 0.026  & 0.231 & 0.942     \\
SS & $\alpha_1$     & 0.004  & 0.277 & 0.964     \\
   & $\beta_1$    & -0.005 & 0.530 & 0.979     \\
   & $\alpha_0$    & -0.004 & 0.333 & 0.959     \\
   & $\beta_0$    & 0.004  & 0.783 & 0.972     \\ \hline
\end{tabular}
\caption{$n=538$ with 3000 simulations. Target coverage probability is 0.95.}
\label{table1}
\end{table}

Table \ref{table1} reports the results for the bias, standard errors, and coverage probability for 3000 replications. The target coverage probability is 0.95. As we observe from the first column, our estimators are unbiased. Our estimators perform well in terms of coverage probabilities as well.

%=============================================================================%
%s_application
%=============================================================================%

\section{Application}\label{sec:application}
\subsection{Background and Data}
Malaria is a life-threatening infectious disease responsible for approximately 1-3 million deaths per year.
Most of these deaths are in children less than five years of age in rural sub-Saharan Africa. The use of insecticide-treated nets (ITNs) has been shown to be a cost-effective way to control malaria. However, the rate of adoption remains low and many households exhibit low willingness to pay (WTP) for ITNs. In addition, positive health externalities generated from using ITNs render the private adoption level that is less than the socially optimal one. For these reasons, public subsidy programs have been proposed to achieve socially optimal coverage rate.

While it has been shown that distributing ITNs for free or at highly subsidized prices is effective in increasing the adoption in the short run, there have been concerns that the short-run, one-time subsidies would lower household's WTPs for the product later, and thus reduce the adoption rate in the long-run. This could happen, for instance, when there exist reference dependence effects in which households anchor their WTPs to previously paid subsidized prices. Consequently, households may be unwilling to pay a higher price for the product later once the subsidies end. 

On the other hand, some argue that short-run subsidies would be beneficial for the long-run adoption since households could learn the benefits of the product better with prior experience. Such learning effects would increase consumer's future WTPs. Moreover, the adoption process can be facilitated with social learning effects in which households learn benefits of the product from their neighbors' prior experiences. As a result, one-time subsidies would also be beneficial for long-run adoption rate and household's WTP.

Since ITNs need to be regularly replaced and re-purchased, understanding the factors determining the short-run and long-run adoption decision is an important task for sustainable public subsidy schemes. Depending on whether reference dependence or learning effects exist, the subsidy schemes would lead to different predictions on the short run and long run demand for ITNs. In this application, therefore, we study the factors affecting the short-run and long-run adoption (purchase) decision of ITNs. In doing so, we allow for possible spillover effects in both short-run and long-run adoption decision. As \cite{dupas} showed, social interactions seem to play an important role in household’s bednet purchase decision. Depending on whether there exist positive or negative peer effects in the short run and in the long run, subsidy effectiveness may vary greatly. 

%--------------------------------------------------------------------------------------------------------------------------------------%
\begin{table}[]
\centering
\begin{tabular}{lllll}
\hline
variable     & definition                   & \multicolumn{1}{l}{mean} & \multicolumn{1}{l}{min} & \multicolumn{1}{l}{max} \\ \hline
degree       & number of neighbors                   & 16.41                    & 1.00                    & 38.00                   \\
$Z$            & 1(high subsidy)              & 0.27                     & 0.00                    & 1.00                    \\
$D$            & 1(adoption at phase 1)                  & 0.47                     & 0.00                    & 1.00                    \\
$Y$            & 1(adoption at phase 2)                  & 0.16                     & 0.00                    & 1.00                    \\
female\_educ & years of educ of female head               & 5.37                     & 0.00                    & 22.00                   \\
wealth       & wealth level                 & 20367.00                 & 0.00                    & 112273.00               \\ \hline
\end{tabular}
\caption{summary statistics ($n=583$)}
\label{table2}
\end{table}

\paragraph{Design of Experiment}
We use data from a two-stage randomized pricing experiment conducted in Kenya by \cite{dupas}. In Phase 1, households within six villages were given a voucher for the bednet at the randomly assigned subsidy level varying from $100\%$ to $40\%$ with the corresponding prices varying from 0 to 250 Ksh. In Phase 2, a year later, all study households in four villages were given a second voucher for a bednet. This time, however, all households faced the same subsidy level of $36\%$.

\paragraph{Data} Let $Z_i$ be a binary indicator representing that household $i$ received a high subsidy (defined as the assigned price less than Ksh 50) in Phase 1. Treatment variable $D_i$ equals to 1 if $i$ purchased a bednet in Phase 1. $Y_i$ is also binary taking value 1 if $i$ purchased a bednet in Phase 2. Following \cite{dupas}, we may interpret $Y_i$ as a proxy for $i$'s WTP for the future bednet.

\textbf{Network} \ Using GPS data, we construct the binarized spatial network. Two households $i$ and $j$ are considered connected (i.e., $G_{ij}=1$) if they live within 500-meter radius. We also consider 250-m, and 750-m radius. Since the results do not differ much, we only report results for 500-m radius.

\textbf{Other Covariates} \ For household pre-treatment covariates, we consider wealth, and the education level of the female head.

Summary statistics of the variables can be found on the Table \ref{table2}. After deleting $25$ isolated nodes, we have $n=538$ observations from four villages.

\begin{figure}[h]
\includegraphics[scale=0.35]{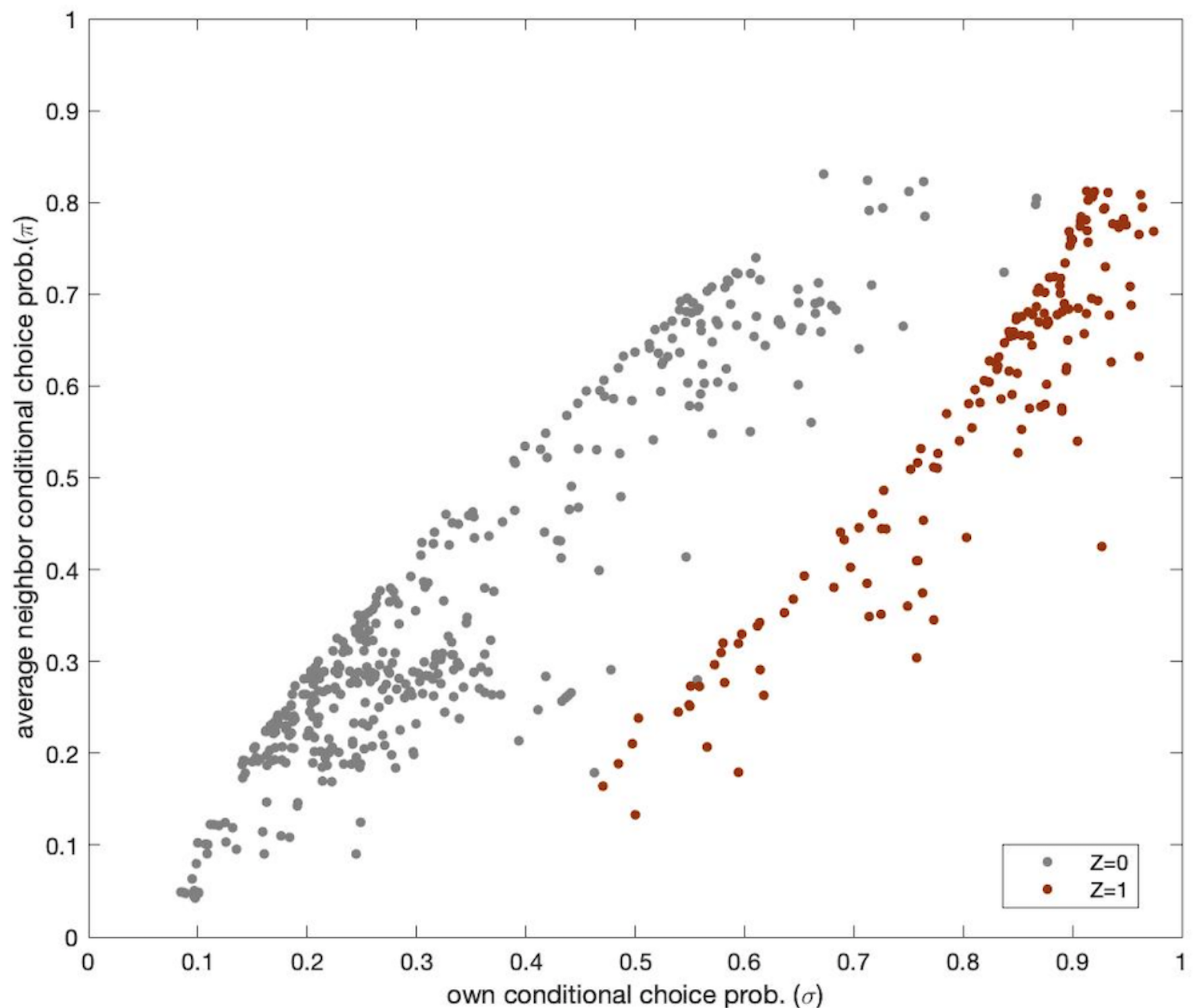}
\centering
\caption{plot of estimated $(\sigma_i^*,\pi_i^*)$}
 \label{fig1}
\end{figure}

%--------------------------------------------------------------------------------------------------------------------------------------%
\subsection{Estimation Results}

\begin{table}[]
\centering
\begin{tabular}{lccc}
\hline
variable    & estimates & marginal effects & p-value \\ \hline
spillover ($\pi$)   & 2.308 & 0.661    & 0.000 \\
subsidy     & 0.694 & 0.199    & 0.000 \\
female-educ & 0.223 & 0.064    & 0.026 \\
wealth      & 0.005 & 0.001    & 0.001 \\ \hline
\end{tabular}
\caption{estimation results for FS model  ($n=583$)}
\label{table3}
\end{table}

\paragraph{Results on the short-run adoption}

We first estimate the equation for the short-run adoption decision using our game-theoretic model. Table \ref{table3} displays the estimates of coefficients, marginal effects\footnote{Marginal effects are computed as the sample average of conditional effects. For instance, the marginal effect of $Z_i$ is computed as $\frac{1}{n}\sum_{i=1}^n\phi(X_i'\hat\theta_1+\hat\theta_2 Z_i+\hat\theta_3\pi_i^*(S,\hat\theta))\hat\theta_2 $.}, as well as associated standard errors and p-values. As anticipated, high-subsidy level is associated with higher adoption of the bednet. Education and wealth are also positively associated with adoption decision in the short run. These variables are all significant at 1 percent level.  Figure \ref{fig1} shows the estimated plot of $(\hat\sigma_i^*,\hat\pi_i^*)$ by the value of $Z_i$. The plot shows clearly that individual $Z_i$ is relevant for the treatment choice.

Our results show strong evidence of the existence of positive spillover effects in the short-run adoption decision. When the average adoption probability of neighbors ($\pi_i^*$) increases by 10 percentage points, $i$’s short-run adoption probability ($\sigma_i^*$) increases by 6.6 percentage points. The resulting conformity effects implies that if we ignore spillover effects in the specification, we would underestimate the full effect of the programs.

%--------------------------------------------------------------------------------------------------------------------------------------%
\begin{table}[]
\centering
\begin{tabular}{lcc|lcc}
\hline
$D=1$            & estimates & p-value & $D=0$            & estimates & p-value \\ \hline
cons             & 0.497     & 0.043   & cons             & 0.128     & 0.174   \\
female-educ    & -0.094    & 0.530   & female-educ    & -0.070    & 0.519   \\
wealth         & 0.003     & 0.388   & wealth         & -0.002    & 0.325   \\
lambda         & 0.059    & 0.767   & lambda         & 0.036    & 0.841   \\
  $\pi$           & -0.347    & 0.324   & $\pi$           & -0.021    & 0.940   \\
  $\pi$*female-educ & 0.031     & 0.906   & $\pi$*female-educ & 0.176     & 0.513   \\
$\pi$*wealth      & -0.003    & 0.610   & $\pi$*wealth      & 0.013     & 0.098   \\
$\pi$*lambda      & 0.317    & 0.375   & $\pi$*lambda      & -0.063     & 0.832   \\ \hline
\end{tabular}
\caption{estimation results for SS model  ($n=583$)}
\label{table4}
\end{table}

\paragraph{Results on the long-run adoption} Table \ref{table4} presents the estimates of own short-run adoption experience ($D_i$) and average adoption probability of neighbors ($\pi_i^*$) on the long-run adoption decision. Unfortunately, we have very limited statistical power except for few constants due to small sample size. However, in terms of magnitudes, estimated coefficients have implications on the spillover effects in the long-run adoption decision.

Using the formula \ref{eq-Y} and \ref{eq-Y2}, we get the following estimated mean response functions:
\begin{eqnarray}\label{ASF}
\widehat{\E}[Y_i(1,\pi)]=0.497-0.347\pi,\quad \widehat{\E}[Y_i(0,\pi)]=0.128-0.02\pi 
\end{eqnarray}

First, let us consider $\widehat{\E}[Y_i(1,\pi)]$. Although the coefficient on $\pi$ is not significant, we observe considerable negative spillover effects in terms of magnitude: If $\pi$ increases by 10 percentage points, the probability of the second-period adoption probability decreases by 3.4 percentage points. This is 
contrary to the positive spillovers observed in the first period adoption decision.  \footnote{\cite{dupas} also report similar results from their reduced-form regression models. Their results show that the adoption in Phase 2 is negatively affected by the share of neighbors who received a high subsidy in Phase 1.} One possible explanation for such negative spillovers in the treated response is that they result from positive health spillovers occurring over time. For instance, household with higher value of $\pi$ would anticipate higher coverage rate in their area, which would result in lower malaria prevalence in the long run. This might make households less likely to re-invest the product later. Such results highlight the importance of distinguishing the mechanism of static spillovers from that of dynamic spillovers.

Such effects do not seem to apply to the untreated households as $\widehat{\E}[Y_i(0,\pi)]$ shows. However, the statistical power is very limited. 

\paragraph{Average Direct Effect} From \ref{ASF}, the average direct effect (ADE) of own short-run adoption on the long-run adoption is computed as follows:
\begin{eqnarray}\label{ADE}
\widehat{\E}[Y_i(1,\pi)-Y_i(0,\pi)]=0.369-0.326\pi 
\end{eqnarray}
The result suggests that the values of ADE vary greatly depending on the value of $\pi$: when $\pi=0$, treated households are 36.9 percentage points more likely to invest in the second bednet. However, such effect declines with the neighborhood exposure rate $\pi$. When $\pi=1$, the effect is almost zero. The fact that $ADE$ is positive for all possible values of $\pi$ points to the existence of learning effects from prior experience, rather than reference dependence effects.

\paragraph{Bias from ignoring spillovers} Suppose that we falsely ignore spillover effects in responses. Using the conventional Heckit model, we obtain the following estimated average treatment effect (ATE): 

$$\hat\E[Y_i(1)-Y_i(0)]=0.038. $$
Above result suggests that the effect of $D$ on $Y$ is very limited. However as equation \ref{ADE} shows, there is substantial heterogeneity in the effect of $D$ on $Y$ depending on values of $\pi$: the effect of $D$ varies from almost 0 percent to 37 percent. Thus, by ignoring the spillover effects, we would draw a misleading conclusion that there is no treatment effect.

\paragraph{Observed heterogeneity in effects} Let us turn to the effect heterogeneity due to observable covariates, education and wealth. For the treated, the effect of education and wealth on the adoption rate seems to be trivial in magnitude: coefficients are close to zero and their associated p-values are large. We also compute the estimates without covariates. The magnitude of the estimates resembles that with covariates. Therefore we do not report the result here. This also suggests that there seems to be little observed heterogeneity in $\E[Y_i(1,\pi)]$ in terms of education and wealth.

On the other hand, for $D_i=0$ case, the magnitudes of the estimates on the covariates are much higher than those for $D_i=1$ case. Consider education first. The interaction between $\pi$ and education suggests that higher education is associated with higher spillover effect — one more year of education increases the effect of $\pi$ from $-0.02$ to $-0.02+0.17=0.15$. Similarly if wealth level increases by 1000 units, the effect on $\pi$ increases by $1.2$ percentage point which is significant at 10 percent. Such results suggest that control households with higher education and higher wealth receive higher positive spillover effect. 

%--------------------------------------------------------------------------------------------------------------------------------------%
%--------------------------------------------------------------------------------------------------------------------------------------%

%--------------------------------------------------------------------------------------------------------------------------------------
%--------------------------------------------------------------------------------------------------------------------------------------%

\subsection{Impact of Counterfactual Policies}
One advantage of our structural approach is that it allows researchers to simulate counterfactual policies. Suppose that a policy-maker is interested in implementing  means-tested subsidy schemes where $Z$ is determined according to the following rule:
\begin{eqnarray}\label{rule1}
Z_i=\1\{wealth_i\le \tau\},\quad\forall i\in \mathcal{N}_n \l
\end{eqnarray}
i.e., household $i$ gets high subsidy only when their wealth level is below some specified threshold $\tau$. The question is: what would be the expected outcome under this new, counterfactual subsidy rule?

This problem is related to the literature on the policy-relevant treatment effects (PRTE: \cite{PRTE}). In this framework, 
each intervention or policy is defined by a manipulation on the exogenous variable $S=(G,X,Z)$. In our setup, we assume that a policy maker has no means of changing the underlying network structure $G$ or pre-treatment covariates $X$. Thus, the only way to change $S$ is through changing $Z$. Let us denote the new counterfactual policy as $S^{new}=(G,X,Z^{new})$  where we set the value of $Z$ as $Z=Z^{new}$, which is not in the data. $i$'s expected outcome under the new policy is given as $\E[Y_i|S=S^{new}]$. Note that for any $S$,
\begin{eqnarray}\label{PRTE}
\E[Y_i|S]=\E[Y_i|D_i=1,S]\Pr(D_i=1|S)+\E[Y_i|D_i=0,S]\Pr(D_i=0|S)
\end{eqnarray}
Under our control function specification, $\E[Y_i|S]$ can be written as follows::
\begin{eqnarray*}
\E[Y_i|S]&=&\sigma_i^*(S)\Big[X_i'\alpha_1+\lambda_1(\sigma_i^*(S))+\Big\{X_i'\beta_1+\lambda_1(\sigma_i^*(S)) \Big\}\pi_i^*(S)\Big]\\
&&+(1-\sigma_i^*(S))\Big[X_i'\alpha_0+\lambda_0(\sigma_i^*(S))+\Big\{X_i'\beta_0+\lambda_0(\sigma_i^*(S)) \Big\}\pi_i^*(S)\Big]\\
&=&\E[Y_i|X_i,\sigma_i^*(S),\pi_i^*(S)]
\end{eqnarray*}
Note that $\E[Y_i|S]$ is a function of $S$ only through $(X_i,\sigma_i^*(S),\pi_i^*(S))$, thus we write $\E[Y_i|X_i,\sigma_i^*(S),\pi_i^*(S)]$. $i$'s expected outcome under new policy is then given by $\E[Y_i|X_i,\sigma_i^*(S^{new}),\pi_i^*(S^{new})]$.

 To estimate this, we first need to compute the new equilibrium choice probabilities: $\{\sigma_i^*(G,X,Z^{new})\}_{i\in \mathcal{N}_n}$ where $Z^{new}$ is determined according to \ref{rule1}. Under the identified first-stage parameters, this is done by solving the new fixed point of the best-response functions under the new data set $S^{new}=(G,X,Z^{new})$. We then estimate $\hat Y_i\equiv \hat\E[Y_i|X_i,\sigma_i^*(S^{new}),\pi_i^*(S^{new})]$ for each $i\in N_n$ using the formula above. Overall impact of policy $S^{new}$ is computed by $\sum_{i=1}^n \hat Y_i/n$.

\paragraph{Results}

\begin{figure}[h]
\includegraphics[width=\textwidth]{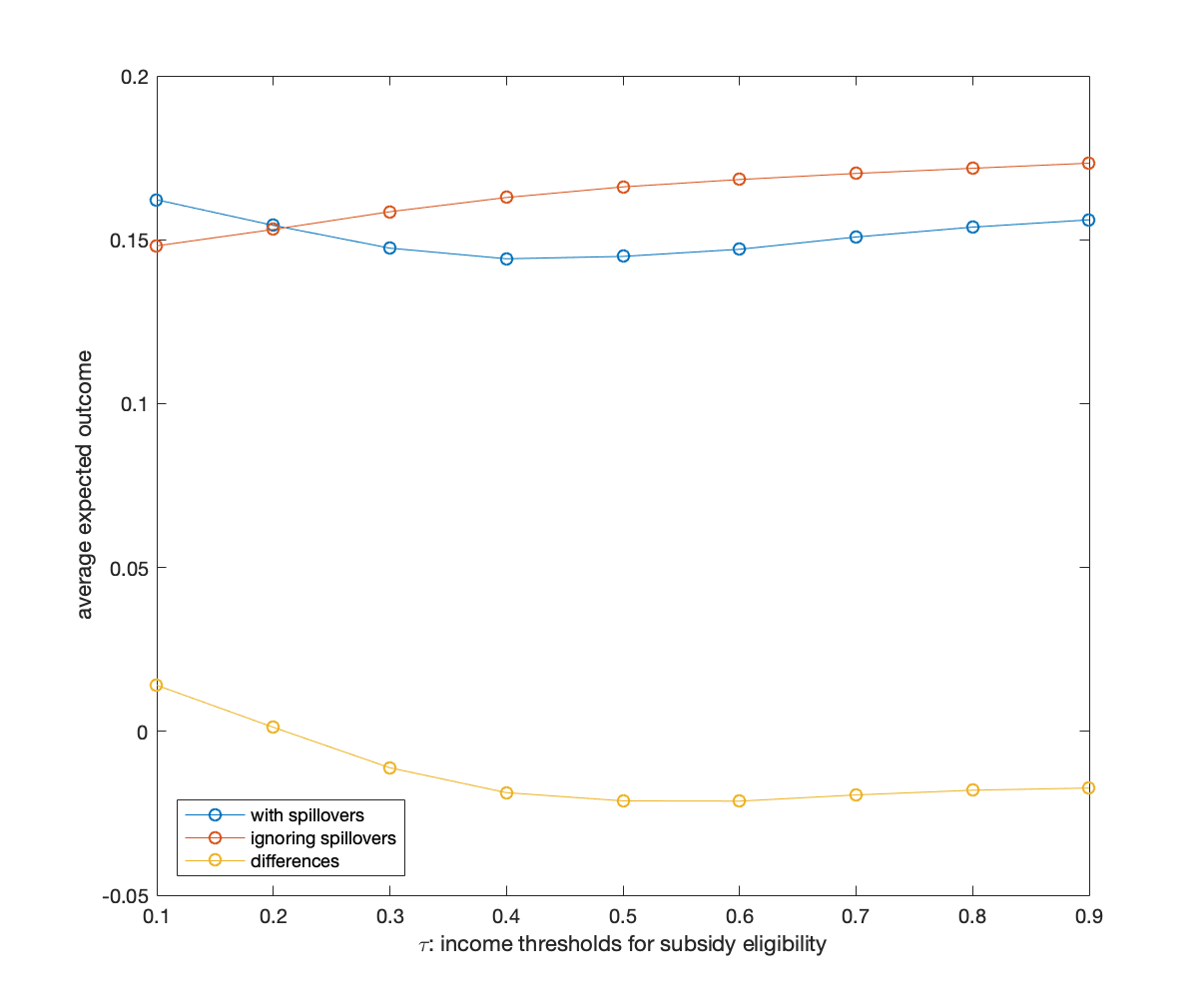}
\centering
\caption{counterfactual impact of means-tested subsidy on LR-adoption}
 \label{fig2}
\end{figure}

See \ref{fig2}. The red line shows the effect of $\tau$ on the overall long-run adoption level when we ignore interference effects. In such case, as $\tau$ increases, the long-run adoption level increases monotonically. This is because as $\tau$ increases, more households get subsidy, and without interference, treated agents are more likely to adopt in the long-run.

In the presence of spillovers, the effect of $\tau$ does not increase monotonically anymore as the blue line shows. Higher $\tau$ also induces higher $\pi_i^*$ which affect long-run adoption negatively. Therefore a priori, we cannot expect that higher $\tau$ would give higher overall long-run adoption rate in the population. In fact, as the blue line shows,
the highest long-run adoption rate is achieved under the subsidy scheme targeting the very lowest percentile households.

 The result also highlights complication involved in the use of subsidies to increase long-run adoption rate. As the result shows, the highest expected coverage is only 17 percent.

%=============================================================================%
%s_conclusion
%=============================================================================%

 \section{Concluding Remarks}
 In this paper, we propose a new methodological framework to analyze randomized experiments with spillovers and noncompliance in a general network setup. Using a game-theoretic framework, we allow for spillover effects to occur at two stages: at the choice stage and outcome stage. Potential outcomes are modeled as a random coefficient model to account for general unobserved heterogeneity. We extend the traditional control function estimator of \cite{heckman_sel} to incorporate spillovers. Finally, we illustrate our methods using \cite{dupas} data and show that our model can be used to evaluate the counterfactual policies. 

In our treatment choice games, we assumed that private information is independently distributed across agents. Relaxing this assumption to allow for network dependence in private information would be a rewarding task. Another important issue is multiple equilibria -- formalizing a problem of policy evaluation and counterfactual prediction in the presence of multiple equilibria is important for realistic policy design. Finally, we conclude by noting that our model can be used to derive an ex ante optimal treatment assignment rule under interference, especially in settings where a social planner should take possible noncompliance and spillover into account.

% observational studies에도 적용이 된다

%--------------------------------------------------------------------------------------------------------------------------------------%
\newpage

\begin{appendix}
\section*{Appendix}
\section{Proof of Theorem \ref{ue}}\label{uniq} Following \cite{xu18}, we show this by contradiction. Define $\bar\sigma_i=\sum_{j\in\mathcal{N}_i}\sigma_j/|\mathcal{N}_i|$. Let $\Gamma(X_i,Z_i,\bar\sigma_i,\theta)=\Phi(X_i'\theta_1+\theta_2 Z_i+\theta_3\bar\sigma_{i})$ be $i$'s best-response function to inputs $(X_i,Z_i, \bar\sigma_i)$, and parameter value $\theta$. Suppose there are two non-identically equilibria $\sigma^*=(\sigma_i^*)_{i\in\mathcal{N}_n}$ and $\sigma^{+}=(\sigma_i^+)_{i\in\mathcal{N}_n}$. By definition, they should satisfy
$$\sigma_i^*=\Gamma(X_i,Z_i,\bar\sigma_{i}^*,\theta),\quad \forall i\in \mathcal{N}_n $$
and
$$\sigma_i^+=\Gamma(X_i,Z_i,\bar\sigma_{i}^+,\theta),\quad \forall i\in \mathcal{N}_n.$$
Taking difference and applying mean-value theorem, we have
\begin{eqnarray*}
\sigma_i^*-\sigma_i^+&=&\Gamma(X_i,Z_i,\bar\sigma_{i}^*,\theta)- \Gamma(X_i,Z_i,\bar\sigma_{i}^+,\theta)\\
&=&\frac{\partial\Gamma(X_i,Z_i,\bar\sigma_i^m,\theta)}{\partial \bar\sigma_i}(\bar\sigma_i^*-\bar\sigma_i^+)
\end{eqnarray*}
where $\bar\sigma_i^m$ is a mean value between $\bar\sigma_i^*$ and $\bar\sigma_i^+$. Taking an absolute value to the LHS,
\begin{eqnarray}
|\sigma_i^{*}-\sigma_i^{+}|&\le& \Big|\frac{\partial \Gamma(X_i,Z_i,\bar\sigma_i^m,\theta)}{\partial\bar\sigma_i}\Big|\cdot |\bar\sigma_i^*-\bar\sigma_i^+|\\
&\le& \Big|\frac{\partial \Gamma(X_i,Z_i,\bar\sigma_i^m,\theta)}{\partial\bar\sigma_i}\Big|\cdot \max_{j\in\mathcal{N}_i}|\sigma_j^*-\sigma_j^+|. \label{124u}
\end{eqnarray}
From the definition of $\Gamma(\cdot)$, observe that
\begin{eqnarray*}
\frac{\partial\Gamma(X_i,Z_i,\bar\sigma_i,\theta)}{\partial \bar\sigma_i}=\frac{\Phi(X_i'\theta_1+\theta_2 Z_i+\theta_3\bar\sigma_i)}{\partial\bar\sigma_i}=\phi\big(X_i'\theta_1+\theta_2 Z_i+\theta_3\bar\sigma_i\big)\theta_3.
\end{eqnarray*}
Thus,
\begin{eqnarray}
\Big|\frac{\partial \Gamma(X_i,Z_i,\bar\sigma_i^m,\theta)}{\partial\bar\sigma_i}\Big|\le|\theta_3|\sup_u\phi(u)\equiv \lambda.\label{UB}
\end{eqnarray}
Therefore we can write \ref{124u} as
$$ |\sigma_i^{*}-\sigma_i^{+}|\le \lambda\max_{j\in\mathcal{N}_i}|\sigma_j^*-\sigma_j^+|.$$

Taking $\max_{i\in N_n}$ to both sides gives,
$$\max_{i\in \mathcal{N}_n}|\sigma_i^{*}-\sigma_i^{+}|\le \lambda \max_{i\in\mathcal{N}_n}\max_{j \in\mathcal{N}_i}|\sigma_j^{*}-\sigma_j^{+}| \le \lambda\max_{k \in \mathcal{N}_n}|\sigma_k^{*}-\sigma_k^{+}| $$
which leads to contradiction when $\lambda<1$. $\blacksquare$

%--------------------------------------------------------------------------------------------------------------------------------------%
\section{Proofs for Asymptotic Results}
\subsection{Proof of consistency of first-stage estimators}\label{thm_consistency_FS}
Let $l_i(\theta)\equiv D_i\ln \sigma_i^*(S,\theta)+(1-D_i)\ln (1-\sigma_i^*(S,\theta))$ be an individual log-likelihood function of $i$. Then $\widehat{\mathcal{L}}_n(\theta)=\frac{1}{n}\sum_{i=1}^n l_i(\theta)$. 

Define 
$$\mathcal{L}_n(\theta)=\E[\widehat{\mathcal{L}}_n(\theta)|S] $$
where the population objective function, $\mathcal{L}_n(\theta)$, depends on $n$ through the public state $S=(G,X,Z)$. Recall that the true parameter is denoted by $\theta^0$.
%--------------------------------------------------------------------------------------------------------------------------------------%
Following \cite{white} Theorem 3.3, we establish consistency result by showing identifiable uniqueness and uniform convergence result.
\paragraph{Identifiable Uniqueness} We show that $\liminf_{n\to\infty} (\mathcal{L}_n(\theta^0)-\mathcal{L}_n(\theta))>0$ for any $\theta$ such that $|\theta-\theta^0|
\ge\epsilon>0$. 
\begin{eqnarray*}
&&-\liminf_{n\to\infty} (\mathcal{L}_n(\theta)-\mathcal{L}_n(\theta^0))\\
&=&\liminf_{n\to\infty} -\frac{1}{n}\sum_{i=1}^n\E\Big[D_i \ln\frac{\sigma_i^*(S,\theta)}{\sigma_i^*(S,\theta^0)}   +(1-D_i)\ln\frac{1-\sigma_i^*(S,\theta)}{1-\sigma_i^*(S,\theta^0)} \Big|S\Big]\\
&=&\liminf_{n\to\infty} -\frac{1}{n}\sum_{i=1}^n\Big[\sigma_i^*(S,\theta^0) \ln\frac{\sigma_i^*(S,\theta)}{\sigma_i^*(S,\theta^0)}   +(1-\sigma_i^*(S,\theta^0))\ln\frac{1-\sigma_i^*(S,\theta)}{1-\sigma_i^*(S,\theta^0)} \Big|S\Big]\\
&\ge&\liminf_{n\to\infty} -\frac{1}{n}\sum_{i=1}^n \ln\big(\sigma_i^*(S,\theta)+1-\sigma_i^*(S,\theta)\big)=0.
\end{eqnarray*}
The second equality follows from $\E[D_i|S]=\sigma_i^*(S,\theta^0)$ and the last weak inequality is due to Jensen's inequality. To show that the inequality holds strictly, we need to rule out the case of $\liminf_{n\to\infty} (\mathcal{L}_n(\theta^0)-\mathcal{L}_n(\theta))=0$. This happens when for some large enough $n$, $\sigma_i^*(S,\theta)=\sigma_i^*(S,\theta^0)$ for all $i\in \mathcal{N}_n=\{1,2,\cdots,n\}$, i.e., there exists $n$ that delivers observationally equivalent choice probabilities.

 Suppose this is the case. By the fixed point requirement, the following needs to be satisfied for any arbitrary $\theta$, including the true parameter $\theta^0$:
$$\Phi^{-1}(\sigma_i^*(S,\theta))=X_i'\theta_1+\theta_2 Z_i+\theta_3\frac{1}{|\mathcal{N}_i|}\sum_{j\in\mathcal{N}_i}\sigma_j^*(S,\theta),\quad \forall i\in \mathcal{N}_n$$
and
$$\Phi^{-1}(\sigma_i^*(S,\theta^0))=X_i'\theta_1^0+\theta_2^0 Z_i+\theta_3^0\frac{1}{|\mathcal{N}_i|}\sum_{j\in\mathcal{N}_i}\sigma_j^*(S,\theta^0),\quad \forall i\in \mathcal{N}_n.$$
If $\sigma_i^*(S,\theta)=\sigma_i^*(S,\theta^0),\ \forall i\in \mathcal{N}_n$, we have,
$$X_i'(\theta_1-\theta_1^0)+Z_i(\theta_2-\theta_2^0)+(\theta_3-\theta_3^0)\frac{1}{|\mathcal{N}_i|}\sum_{j\in\mathcal{N}_i}\sigma_j^*(S,\theta^0)=0,\quad\forall i\in \mathcal{N}_n. $$
Equivalently, $R_i'(\theta-\theta^0)=0,\ \forall i\in \mathcal{N}_n$ where $R_i$ is defined as in  Theorem \ref{thm:A}. It follows that $(\theta-\theta^0)'\sum_{i=1}^n R_iR_i'(\theta-\theta^0)=0$. Given the assumption that $\sum_{i=1}^n R_iR_i'$ is positive definite for all large enough $n$, above equation holds only under $\theta=\theta^0$ leading to contradiction. $\square$\ \\

%--------------------------------------------------------------------------------------------------------------------------------------%

Next, we verify that $\sup_{\theta\in\Theta}|\widehat{\mathcal{L}}_n(\theta)-\mathcal{L}_n(\theta)|\xrightarrow{p}   0.$ We first shows the pointwise convergence holds. Uniform convergence follows then from Lipschitz conditions.
\paragraph{Pointwise Convergence} We first show that for any $\theta\in\Theta$, $|\widehat{\mathcal{L}}_n(\theta)-\mathcal{L}_n(\theta)|\xrightarrow{p}   0.$ It can be shown that
\begin{eqnarray*}
\widehat{\mathcal{L}}_n(\theta)-\mathcal{L}_n(\theta)
=\frac{1}{n}\sum_{i=1}^n\Big\{ \underbrace{(D_i-\sigma_i^*(S,\theta^0))\ln\frac{\sigma_i^*(S,\theta)}{1-\sigma_i^*(S,\theta)}}_{\zeta_i}
\Big\}.
\end{eqnarray*}
$\{\zeta_i\}_{i=1}^n$ is conditionally independent with mean zero given $S$. It is also uniformly bounded due to Lemma \ref{unif_bdd}. Therefore we can apply a LLN for independent observations (e.g., Markov) and the result follows.

%--------------------------------------------------------------------------------------------------------------------------------------%

\paragraph{Uniform Convergence} Given pointwise convergence result, uniform convergence follows if we can establish that $\{\widehat{\mathcal{L}}_n(\theta)-\mathcal{L}_n(\theta)\}_n$ is stochastically equicontinuous on $\Theta$ (theorem 1 in \cite{andrews}). Sufficient condition for this is to show that the summand in the sample objective function $\{l_i(\theta)\}$ is Lipschitz
(Assumption W-LIP in \cite{andrews}). Note that
\begin{eqnarray*}
\nabla_\theta l_i(\theta)= D_i\frac{\nabla_\theta \sigma_i^*(S,\theta)}{\sigma_i^*(S,\theta)}+(1-D_i)\frac{-\nabla_\theta\sigma_i^*(S,\theta)}{1-\sigma_i^*(S,\theta)}
\end{eqnarray*}
which is bounded by
\begin{eqnarray*}
|\nabla_\theta l_i(\theta)|\le \Big|\frac{\nabla_\theta\sigma_i^*(S,\theta)}{\sigma_i^*(S,\theta)}\Big|+\Big|\frac{\nabla_\theta\sigma_i^*(S,\theta)}{1-\sigma_i^*(S,\theta)}\Big|.
\end{eqnarray*}
By Lemma \ref{unif_bdd} and Lemma \ref{unif_bdd2}, $\sigma_i^*(S,\theta)$ and $\nabla_\theta\sigma_i^*(S,\theta)$ are uniformly bounded. Therefore $\{l_i(\theta)\}$ is Lipschitz-continuous and the result follows. $\blacksquare$

%=============================================================================%

\subsection{Proof of asymptotic normality of first-stage estimators}\label{thm_norm_FS}

%---------------------------------------------------------------------------------------------------------------------------------------%
$\hat\theta$ should satisfy the first-order condition for maximization: $\nabla_\theta\widehat{\mathcal{L}}_n(\hat\theta)=0$. Given that $\widehat{\mathcal{L}}_n(\hat\theta)$ is smooth, we can apply the mean-value theorem to the first-order condition around the true parameter $\theta^0$:
\begin{eqnarray}
&&\nabla_\theta\widehat{\mathcal{L}}_n(\hat\theta)=\nabla_\theta\widehat{\mathcal{L}}_n(\theta^0)+\nabla_{\theta\theta}\widehat{\mathcal{L}}_n(\bar\theta)(\hat\theta-\theta^0)=0\\
&\Longleftrightarrow& \sqrt{n}(\hat\theta-\theta^0)=-(\nabla_\theta\widehat{\mathcal{L}}_n(\bar\theta))^{-1}\sqrt{n}\nabla_\theta\widehat{\mathcal{L}}_n(\hat\theta^0) \label{FO}
\end{eqnarray}
where $\bar\theta$ is a mean value of the line joining $\hat\theta$ and $\theta^0$. 
%---------------------------------------------------------------------------------------------------------------------------------------%
 Define the Hessian matrix as
$$\mathcal{H}_n(\theta)=\E\Big[\frac{1}{n}\sum_{i=1}^n\nabla_{\theta\theta} l_i(\theta)\Big|S\Big] $$
and the information matrix as 
$$\mathcal{I}_n(\theta)=\E\Big[\frac{1}{n}\sum_{i=1}^n\nabla_{\theta} l_i(\theta)\nabla_{\theta} l_i(\theta)'\Big|S\Big]. $$
We first show that $\nabla_{\theta\theta}\widehat{\mathcal{L}}_n(\bar\theta)-\mathcal{H}_n(\theta^0)\xrightarrow{p}   0$ (ULLN of the Hessian matrix) and then $\sqrt{n}\mathcal{I}_n^{-1}(\theta^0)\nabla_\theta\widehat{\mathcal{L}}_n(\hat\theta^0)\xrightarrow{d}   N(0,I_{dim(\theta)})$ (CLT on the score).

%---------------------------------------------------------------------------------------------------------------------------------------%
\paragraph{ULLN of the Hessian Matrix}\label{hess} We show that $\nabla_{\theta\theta} \widehat{\mathcal{L}}_n(\bar\theta)-\mathcal{H}_n(\theta^0)\xrightarrow{p}   0$. Note that
\begin{eqnarray*}
&&\nabla_{\theta\theta} \widehat{\mathcal{L}}_n(\bar\theta)-\mathcal{H}_n(\theta^0)\\
&&=
\underbrace{\frac{1}{n}\sum_{i=1}^n\nabla_{\theta\theta} l_i(\bar\theta)-\frac{1}{n}\sum_{i=1}^n\nabla_{\theta\theta} l_i(\theta^0)}_{A}+\underbrace{\frac{1}{n}\sum_{i=1}^n\nabla_{\theta\theta} l_i(\theta^0)-\E\Big[\frac{1}{n}\sum_{i=1}^n\nabla_{\theta\theta} l_i(\theta^0)\Big|S\Big]}_{B}
\end{eqnarray*}
First, $A=o_p(1)$ since $\hat\theta-\theta^0\xrightarrow{p}   0$ and $\nabla_{\theta\theta}l_i(\cdot)$ is continuous as a result of Lemma \ref{unif_bdd3}. Next, note that
\begin{eqnarray*}
B=\frac{1}{n}\sum_{i=1}^n\Big\{\underbrace{\nabla_{\theta\theta}l_i(\theta^0)-\E\big[\nabla_{\theta\theta}l_i(\theta^0) \big|S\big] \Big\}}_{\xi_i}
\end{eqnarray*}
$\{\xi_i\}$ is independent conditional on $S$ with mean zero. Also by Lemma \ref{unif_bdd3}, it is uniformly bounded. Therefore by LLN for independent observations, $B=o_p(1)$. 
%---------------------------------------------------------------------------------------------------------------------------------------%
\paragraph{CLT on the Score}\label{score}
Note that $\sqrt{n}\nabla_\theta\widehat{\mathcal{L}}_n(\theta^0)=\sqrt{n}\frac{1}{n}\sum_{i=1}^n\nabla_\theta  l_i(\theta^0)$ and that $\{\nabla_\theta  l_i(\theta^0)\}$ is independently distributed conditional on $S$ with the uniformly bounded conditional variance $\mathcal{I}_n(\theta^0)$. Therefore we can apply Lyapunov's CLT for independent observations to get $\sqrt{n}\mathcal{I}_n^{-1/2}(\theta^0)\nabla_\theta\widehat{\mathcal{L}}_n(\theta^0)\xrightarrow{d}   N(0,I)$.\ \\

%--------------------------------------------------------------------------------------------------------------------------------------%
Combining all these results, we see that the equation \ref{FO} can be written as
\begin{eqnarray*}
\sqrt{n}(\hat\theta-\theta^0)=-(\mathcal{H}_n(\theta^0)+o_p(1))^{-1}\mathcal{I}_n(\theta^0)^{1/2} \sqrt{n}\mathcal{I}_n(\theta^0)^{-1/2}\nabla\widehat{\mathcal{L}_n}(\theta^0) 
\end{eqnarray*}
By the information matrix inequality, when the model is correctly specified, $\mathcal{H}_n(\theta^0)=-\mathcal{I}_n(\theta^0)$ so that we have
\begin{eqnarray*}
\sqrt{n}(\hat\theta-\theta^0)=(\mathcal{I}_n(\theta^0)+o_p(1))^{-1}\mathcal{I}_n(\theta^0)^{1/2} \sqrt{n}\mathcal{I}_n(\theta^0)^{-1/2}\nabla\widehat{\mathcal{L}_n}(\theta^0) 
\end{eqnarray*}
Under the assumption that $\mathcal{I}_n(\theta^0)$ is nonsingular, we get the desired result:

$$\sqrt{n}(\mathcal{I}_n^{-1}(\theta^0))^{-1/2}(\hat\theta-\theta^0)\xrightarrow{d}   N(0,I_{dim(\theta)}). $$
$\blacksquare$
%--------------------------------------------------------------------------------------------------------------------------------------%
%=============================================================================%

%--------------------------------------------------------------------------------------------------------------------------------------%

\subsection{Proof of consistency of second-stage estimators}\label{thm_consistency}
Our estimators are based on the following moment conditions
\begin{eqnarray*}
\E[Y_i|D_i=1,S]=W_{i}'\gamma_1^0,\quad \E[Y_i|D_i=0,S]=W_i'\gamma_0^0
\end{eqnarray*}
Let us focus on $\hat\gamma_1$ case as $\hat\gamma_0$ case can be analyzed in an analogous way.

Given the moment condition $\E[Y_i|D_i=1,S]=W_{i}'\gamma_1^0$, we write the equation in error form as
$$Y_i=W_i'\gamma_1^0+\epsilon_{1i},\quad \E[\epsilon_{1i}|D_i=1,S]=0.$$
Estimator for $\gamma_1$ is defined as 
\begin{eqnarray}
\hat\gamma_1 &=& \arg\min_{\gamma_1} \frac{1}{n}\sum_{i=1}^n D_i\big(Y_i-\hat W_i'\gamma_1\big)^2 \\
&=& \arg\min_{\gamma_1} \frac{1}{n}\sum_{i=1}^n\big(D_iY_i-D_i\hat W_i'\gamma_1\big)^2\\
&=&\Big\{\sum_{i=1}^n D_i\hat W_i\hat W_i'\Big\}^{-1}\sum_{i=1}^nD_i\hat W_iY_i \label{thetahat}
\end{eqnarray}
Note that $D_iY_i=D_iY_{i}(1,\pi_i^*(S,\theta^0))=D_i(W_{i}'\gamma_1^0+\epsilon_{1i})=D_i\big(\hat W_{i}'\gamma_1^0+\epsilon_{1i}-(\hat W_{i}-W_{i})'\gamma_1^{0}\big)$. Plugging this into \ref{thetahat} gives that
\begin{eqnarray*}
\hat\gamma_1&=&\big(\sum_{i=1}^n D_i\hat W_i\hat W_i'\big)^{-1}\sum_{i=1}^nD_i\hat W_i\big(\hat W_{i}'\gamma_1^0+\epsilon_{1i}-(\hat W_{i}-W_{i})'\gamma_1^0\big)\\
&=&\gamma_1^0+ \big(\sum_{i=1}^n D_i\hat W_i\hat W_i'\big)^{-1}\sum_{i}D_i\hat W_i \big(\epsilon_{1i}-(\hat W_{i}-W_{i})'\gamma_1^0\big)
\end{eqnarray*}
so that
\begin{eqnarray}\label{AB}
\hat\gamma_1-\gamma_1^0=  \big(\underbrace{\frac{1}{n}\sum_{i=1}^n D_i\hat W_i\hat W_i'}_{\textbf{A}}\big)^{-1}\underbrace{\frac{1}{n}\sum_{i}D_i\hat W_i \big(\epsilon_{1i}-(\hat W_{i}-W_{i})'\gamma_1^0\big)}_{\textbf{B}}=\textbf{A}^{-1}\textbf{B}.
\end{eqnarray}
%--------------------------------------------------------------------------------------------------------------------------------------%
\paragraph{Part $\bf{A}$} We show that $\frac{1}{n}\sum_{i=1}D_i\hat W_i\hat W_i'-\E[\frac{1}{n}\sum_{i=1}^n D_iW_iW_i'|S]=o_p(1)$. Decompose $\frac{1}{n}\sum_{i=1}D_i\hat W_i\hat W_i'-\E[\frac{1}{n}\sum_{i=1}^n D_iW_iW_i'|S]$ into two parts as follows:
\begin{eqnarray*}
\underbrace{\frac{1}{n}\sum_{i=1}D_i\hat W_i\hat W_i'-\frac{1}{n}\sum_{i=1}^n D_iW_iW_i'}_{(a)}+\underbrace{\frac{1}{n}\sum_{i=1}^n D_iW_iW_i'-\frac{1}{n}\sum_{i=1}^n\E[D_iW_iW_i'|S]}_{(b)}.
\end{eqnarray*}
$(a)=o_p(1)$ since $\hat\theta-\theta^0\xrightarrow{p} 0$ and $W_i(\theta)$ is continuous in $\theta$. For $(b)$, note that the summand $\{D_iW_iW_i'-\E[D_iW_iW_i'|S]\}$ is conditionally independent given $S$ with mean zero. It is also uniformly bounded. Therefore by LLN, $(b)=o_p(1)$.  Finally, invertibility of $\E[\frac{1}{n}\sum_{i=1}^n D_iW_iW_i'|S]$ follows from the identification condition.
%--------------------------------------------------------------------------------------------------------------------------------------%
\paragraph{Part $\bf{B}$} Since $\hat W_i-W_i=o_p(1)$, we can write it $\bf{B}$ as 
\begin{eqnarray*}
\frac{1}{n}\sum_{i=1}^n D_i(W_i+o_p(1))(\epsilon_{1i}-o_p(1)) = \frac{1}{n}\sum_{i=1}^n D_i W_i\epsilon_{1i}
\end{eqnarray*}
Similar argument as above shows that
$$\frac{1}{n}\sum_{i=1}^n\Big(D_iW_i\epsilon_{1i}-\E[D_iW_i\epsilon_{1i}|S]\Big)=o_p(1).$$
It follows from the moment condition $\E[\epsilon_{1i}|D_i=1,S]=0$ that $\E[D_iW_i\epsilon_{1i}|S]=0$. Therefore we conclude that 
$$\textbf{B}=\frac{1}{n}\sum_{i=1}^n D_iW_i\epsilon_{1i} + o_p(1) = o_p(1).$$
Combining with the result on part $\bf{A}$, we conclude that $\hat\gamma_1-\gamma_1=o_p(1)$. $\blacksquare$
%=============================================================================%
\subsection{Proof of asymptotic normality of second-stage estimators}\label{thm_norm_SS}
From \ref{AB}, 
\begin{eqnarray}
\sqrt{n}(\hat\gamma_1-\gamma_1^0)=\Big(\frac{1}{n}\sum_{i=1}^n D_i\hat W_i\hat W_i'\Big)^{-1}\frac{1}{\sqrt{n}}\sum_{i}D_i\hat W_i \Big(\epsilon_{1i}-\gamma_1^{0\prime}(\hat W_{i}-W_{i})\Big)\\
=\Big(\E[\frac{1}{n}\sum_{i=1}^n D_iW_iW_i'|S]+o_p(1)\Big)^{-1}\underbrace{\frac{1}{\sqrt{n}}\sum_{i}D_i\hat W_i \Big(\epsilon_{1i}-\gamma_1^{0\prime}(\hat W_{i}-W_{i})\Big)}_{\bf{C}}\label{AC}
\end{eqnarray}
where the last step has been established in the previous section. Consider the term $\hat W_i-W_i$ in $\bf{C}$. By mean-value theorem,
\begin{eqnarray*}
&&\hat W_i-W_i=W_i(\hat\gamma_1)-W_i(\gamma_1^0)=\nabla_{\gamma_1} W_i(\bar\gamma_1)(\hat\gamma_1-\gamma_1^0) \\
&\Longrightarrow&\sqrt{n}(\hat W_i-W_i)=\nabla_{\gamma_1} W_i(\bar\gamma_1)\sqrt{n}(\hat\gamma_1-\gamma_1^0) 
\end{eqnarray*}
where $\bar\gamma_1$ is a mean value of the line joining $\hat\gamma_1$ and $\gamma_1^0$. By the asymptotic normality of the first-step estimator $\hat\theta$ as in the equation \ref{FSnorm}, we can show that $\sqrt{n}(\hat\theta-\theta^0)$ is asymptotically linear. Specifically, define the influence function as $\eta_i=\E[\frac{1}{n}\sum_{i=1}^n\nabla_\theta l_i(\theta^0)\nabla_\theta l_i(\theta^0)'|S]\nabla_\theta l_i(\theta^0)$, then
$$\sqrt{n}(\hat\theta-\theta^0)=\frac{1}{\sqrt{n}}\sum_{i=1}^n\eta_i+o_p(1).$$

Therefore the term $\bf{C}$ in $\sqrt{n}(\hat\gamma_1-\gamma_1^0)$ can be written as
\begin{eqnarray*}
\frac{1}{\sqrt{n}}\sum_{i}D_i\hat W_i (\epsilon_{1i}-\gamma_1^{0\prime} (\hat W_{i}-W_{i}))=\frac{1}{\sqrt{n}}\sum_{i=1}^n D_i\hat W_{i}\epsilon_{1i}
- \frac{1}{n}\sum_{i=1}^n D_i\hat W_i \gamma_1^{0\prime}\sqrt{n}(\hat W_i-W_i)\\
=\underbrace{\frac{1}{\sqrt{n}}\sum_{i=1}^n D_i\hat W_{i}\epsilon_{1i}}_{\textbf{C}(a)}
-\underbrace{\Big\{\frac{1}{n}\sum_{i=1}^n D_i\hat W_i \gamma_1^{0\prime} \nabla_{\gamma_1}W_i(\bar\gamma_1)\Big\} }_{\textbf{C}(b)} \frac{1}{\sqrt{n}}\sum_{i=1}^n\eta_i+o_p(1)
\end{eqnarray*}

We first show that $\bf{C}(a)$ can be replaced by $\frac{1}{\sqrt{n}}\sum_{i=1}^n D_iW_i\epsilon_{1i}$ and that $\bf{C}(b)$ can be replaced by $\E[\frac{1}{n}\sum_{i=1}^n D_iW_i\gamma_1^{0\prime}\nabla_{\gamma_1}W_i(\gamma_1^0)]$. 

%=============================================================================%
\paragraph{Part \bf{C}(a)}
We show that
$$\frac{1}{\sqrt{n}}\sum_{i=1}^n\Big(D_i\hat W_i\epsilon_{1i}-D_iW_i\epsilon_{1i} \Big)\xrightarrow{p}   0$$
Note htat
\begin{eqnarray}
\frac{1}{\sqrt{n}}\sum_{i=1}^n D_i(\hat W_i-W_i)\epsilon_{1i}
&=&\frac{1}{\sqrt{n}}\sum_{i=1}^n D_i \nabla_{\gamma_1} W_i(\bar\gamma_1)(\hat\gamma_1-\gamma_1^0\big) \epsilon_{1i}\\
&=&\frac{1}{n}\sum_{i=1}^n D_i\nabla_{\gamma_1} W_i(\bar\gamma_1)\sqrt{n}(\hat\gamma_1-\gamma_1^0)\epsilon_{1i}\\
&=&\frac{1}{n}\sum_{i=1}^n D_i\nabla_{\gamma_1} W_i(\bar\gamma_1)\big(\frac{1}{\sqrt{n}}\sum_{i=1}^n\eta_i\big)\epsilon_{1i}\\
&=&\Big(\frac{1}{n}\sum_{i=1}^n D_i \nabla_{\gamma_1}W_i(\bar\gamma_1)\epsilon_{1i}\Big)\frac{1}{\sqrt{n}}\sum_{i=1}^n\eta_i \label{ff}
\end{eqnarray}
It can be shown easily that $\frac{1}{n}\sum_{i=1}^n \Big(D_i\nabla_{\gamma_1} W_i(\gamma_1^0)\epsilon_{1i}-\E[D_i\nabla_{\gamma_1} W_i(\bar\gamma_1)\epsilon_{1i}|S] \Big)\xrightarrow{p}   0 $ where $\E[D_i\nabla_{\gamma_1} W_i(\gamma_1^0)\epsilon_{1i}|S]=0$ from the moment condition. Therefore equation \ref{ff} becomes $o_p(1)\times O_p(1)$ and the result follows. $\blacksquare$
%=============================================================================%
\paragraph{Part $\bf{C(b)}$} We show that
$$\frac{1}{n}\sum_{i=1}^n D_i\hat W_i\gamma_1^{0\prime}\nabla_{\gamma_1} W_i(\bar\gamma_1)-\E[\frac{1}{n}\sum_{i=1}^n D_iW_i\gamma_1^{0\prime}\nabla_{\gamma_1}W_i(\gamma_1^0)|S]=o_p(1).$$
Decompose the LHS as
\begin{eqnarray*}
&&\underbrace{\frac{1}{n}\sum_{i=1}^n D_i\hat W_i\gamma_1^{0\prime}\nabla_{\gamma_1} W_i(\bar\gamma_1)
-\frac{1}{n}\sum_{i=1}^n D_iW_i\gamma_1^{0\prime}\nabla_{\gamma_1} W_i(\gamma_1^0)}_{A}\\
&&+\underbrace{\frac{1}{n}\sum_{i=1}^n D_iW_i\gamma_1^{0\prime}\nabla_{\gamma_1} W_i(\gamma_1^0)
-\E[\frac{1}{n}\sum_{i=1}^n D_iW_i\gamma_1^{0\prime}\nabla_{\gamma_1}W_i(\gamma_1^0)|S]}_{B}.
\end{eqnarray*}
$A=o_p(1)$ since  $\hat\theta-\theta^0\xrightarrow{p}   0$. Also, since $\{D_iW_i\gamma_1^{0\prime}\nabla_{\gamma_1}W_i(\gamma_1^0)\}$ are conditionally independent given $S$ and uniformly bounded, we can apply Markov LLN to show that $B=o_p(1)$. $\blacksquare$
%=============================================================================%
\paragraph{Combining all the results,} term $\bf{C}$ can be written as
\begin{eqnarray*}
\bf{C}
&=&\frac{1}{\sqrt{n}}\sum_{i=1}^n D_iW_i\epsilon_{1i}-\E\Big[\frac{1}{n}\sum_{i=1}^n D_iW_i\gamma_1^{0\prime}\nabla_{\gamma_1} W_i(\gamma_1^0)\Big|S\Big]\frac{1}{\sqrt{n}}\sum_{i=1}^n\eta_i+o_p(1)\\
&=&\frac{1}{\sqrt{n}}\sum_{i=1}^n\underbrace{\Big\{ D_iW_i\epsilon_{1i}-\E\Big[\frac{1}{n}\sum_{i=1}^n D_iW_i\gamma_1^{0\prime}\nabla_{\gamma_1}W_i(\gamma_1^0)\Big|S\Big]\eta_i  \Big\}}_{\zeta_i}.
\end{eqnarray*}
Since $\zeta_i|S$ has a mean zero and is independently distributed, we can apply CLT for the independent observation and get $\Psi_n^{-1/2}\frac{1}{\sqrt{n}}\sum_{i=1}^n\zeta_i\xrightarrow{d}   N(0,I_{dim(\gamma_1)})$ where $\Psi_n=\frac{1}{n}\sum_{i=1}^n \E[\zeta_i\zeta_i'|S]$ which can be simplified as 
\begin{eqnarray*}
\frac{1}{n}\sum_{i=1}^n \E[D_i W_iW_i'\epsilon_{1i}^2]+\E\Big[\frac{1}{n}\sum_{i=1}^n D_iW_i\gamma_1^{0\prime}\nabla_{\gamma_1} W_i(\gamma_1^0)\Big|S\Big]\frac{1}{n}\sum_{i=1}^n\E[\eta_i\eta_i'|S]\E\Big[\frac{1}{n}\sum_{i=1}^n D_iW_i\gamma_1^{0\prime}\nabla_{\gamma_1}W_i(\gamma_1^0)\Big|S\Big]'
\end{eqnarray*}

as the cross-terms get crossed out due to $\E[\epsilon_{1i}\eta_i'|S]=0$, i.e., the first- and second-stage moments are uncorrelated. Finally, from \ref{AC}, and by defining $\Upsilon_n=\E[\frac{1}{n}\sum_{i=1}^n W_iW_i'|S]$, we have
$$\Lambda_n^{-1/2}\sqrt{n}(\hat\gamma_1-\gamma_1^0)\xrightarrow{d}   N(0,I_{dim(\gamma_1)}) $$
for $\Lambda_n=\Upsilon_n^{-1}\Psi_n\Upsilon_n^{-1}$ as desired. $\blacksquare$

%--------------------------------------------------------------------------------------------------------------------------------------%
\section{Auxiliary Lemmas}
\begin{lemma}[uniform boundedness of $\sigma_i^*(S,\theta)$]\label{unif_bdd} There exists a constant $C\in(0,1)$ such that $\sigma_i^*(S,\theta)\ge C$ for any $i,S,\theta$ and $n$. 
\end{lemma}
\subparagraph{(Proof)} As in \ref{uniq}, let us define agent's best-response function as $\Gamma(X_i,Z_i,\bar\sigma_i,\theta)=\Phi(X_i'\theta_1+\theta_2 Z_i+\theta_3\bar\sigma_{i})$.
Recall that $\sigma_i^*(S,\theta)=\Phi(X_i'\theta_1+\theta_2 Z_i+\theta_3\pi_i^*(S,\theta))$.The result follows since $X_i$ is bounded, $Z_i$ is binary, and $\pi_i^*(S,\theta)\le 1$,  $\blacksquare$%--------------------------------------------------------------------------------------------------------------------------------------%

\begin{lemma}[uniform boundedness of $\nabla\sigma_i$]\label{unif_bdd2} Suppose $\lambda<1$. There exists a finite constant $C_{1}$ such that
$$\sup_{i,n,S,\theta,k}\big|\frac{\partial\sigma_i^*(S,\theta)}{\partial\theta_k} \big|<C_{1} <\infty.$$
\end{lemma}
\subparagraph{(Proof)} Recall that 
$$\sigma_i^*(S,\theta)=\Gamma(X_i,Z_i,\bar\sigma_i^*(S,\theta),\theta).$$
Differentiating above equation with respect to $\theta_k$ gives
$$\frac{\partial\sigma_i^*(S,\theta)}{\partial\theta_k}=\frac{\partial \Gamma(X_i,Z_i,\bar\sigma_i^*,\theta)}{\partial\theta_k} +\frac{\partial \Gamma(X_i,Z_i,\bar\sigma_i^*,\theta)}{\partial\bar\sigma_i^*}\frac{\partial\bar\sigma_i^*(S,\theta)}{\partial\theta_k}$$
 Equivalently,
\begin{eqnarray}
\frac{\partial\sigma_i^*(S,\theta)}{\partial\theta_k}=\frac{\partial \Gamma(X_i,Z_i,\bar\sigma_i^*,\theta)}{\partial\theta_k} +\frac{1}{|\mathcal{N}_i|}\sum_{j\in\mathcal{N}_i}\frac{\partial \Gamma(X_i,Z_i,\bar\sigma_i^*,\theta)}{\partial\bar\sigma_i^*}\frac{\partial \sigma_j^*(S,\theta)}{\partial\theta_k}\label{implicit}
\end{eqnarray}
which gives the implicit function of $[\partial\sigma_i^*(S,\theta)/\partial\theta_k]_{i\in \mathcal{N}_n}$. Let us write \ref{implicit} in matrix form by defining the following:
\begin{itemize} 
\item Let $\chi_n$ be $n\times 1$ vector with $i$th component $\partial\sigma_i^*(S,\theta)/\partial\theta_k$. 
\item Let $D_n$ be $n\times n$ matrix with $ij$th element
$$\frac{1}{|\mathcal{N}_i|}\frac{\partial\Gamma(X_i,Z_i,\bar\sigma_i^*,\theta)}{\partial\bar\sigma_i^*}$$
if $G_{ij}=1$ and zero if $G_{ij}=0$.
\item Let $\tau_n$ be $n\times 1$ vector with $i$th component $\frac{\partial \Gamma(X_i,Z_i,\bar\sigma_i^*,\theta)}{\partial\theta_k}$.
\end{itemize}

Then we can write the system \ref{implicit} as $\chi_n=D_n\chi_n+\tau_n$ or equivalently,
$$(I_n-D_n)\chi_n=\tau_n $$
which is invertible if $||D_n||_\infty<1$ where the induced matrix norm $||D_n||_\infty$ is the maximum of the absolute values of row sums, i.e.,
$$||D_n||_\infty=\max_{i\in \mathcal{N}_n} \Big|\frac{\partial\Gamma(X_i,Z_i,\bar\sigma_i^*,\theta)}{\partial\bar\sigma_i^*}\Big|.$$ 
 \ref{UB} implies that $||D_n||_{\infty}\le\lambda$, thus $||D_n||_{\infty}<1$. Therefore $D_n$ is invertible and $(I_n-D_n)^{-1}=\sum_{t=0}^\infty D_n^t$. It follows that $\chi_n=(\sum_{t=0}^\infty D_n^t)\tau_n$. Taking sup norm gives
$$||\chi_n||_\infty\le \sum_{t=0}^\infty ||D_n^t||_\infty ||\tau_n||_\infty =\frac{||\tau_n||_\infty}{1-\lambda} <\frac{C_\tau}{1-\lambda}$$
since RHS does not depend on $(i,n,z_n,\theta,k)$, we have the desired result. $\blacksquare$

%--------------------------------------------------------------------------------------------------------------------------------------%

\begin{lemma}[uniform boundedness of $\nabla^2\sigma_i$]\label{unif_bdd3} Suppose $\lambda<1$. There exists a finite constant $C_2$ such that
$$|\frac{\partial^2\sigma_i^*(S,\theta)}{\partial\theta_m\partial\theta_k} |<C_2 <\infty$$
for any $i,n,S,\theta,k,m$ a.s. 
\end{lemma}
\subparagraph{(Proof)}

Fix $m$. Differentiating the equation \ref{implicit} w.r.t. $\theta_m$ gives
\begin{eqnarray*}
\frac{\partial^2 \sigma_i}{\partial\theta_m\partial\theta_k}
=\frac{\partial^2\Gamma}{\partial\theta_m\partial\theta_k}+
\frac{\partial^2\Gamma}{\partial\bar\sigma_i\partial\theta_k}\frac{\partial\bar\sigma_i}{\partial\theta_m}+
\frac{\partial\Gamma}{\partial\bar\sigma_i}\frac{\partial^2\bar\sigma_i}{\partial\theta_m\partial\theta_k}+\frac{\partial\bar\sigma_i}{\partial\theta_k}\Big\{\frac{\partial^2\Gamma}{\partial\bar\sigma_i^2} \frac{\partial\bar\sigma_i}{\partial\theta_m}+\frac{\partial^2\Gamma}{\partial\theta_m\partial\bar\sigma_i} \Big\}.
\end{eqnarray*}
Let us write it compactly as follows:
\begin{eqnarray}
\framebox[1.1\width]{$\partial^2_{mk}\sigma_i$}
=\Gamma_{mk}
+\Gamma_{\bar\sigma k}\partial_m\bar\sigma_i
+\Gamma_{\bar\sigma}\framebox[1.1\width]{$\partial^2_{mk}\bar\sigma_i$}
+\Gamma_{\bar\sigma\bar\sigma}\partial_k\bar\sigma_i\partial_m\bar\sigma_i
+\Gamma_{\bar\sigma m}\partial_k\bar\sigma_i.\label{ODE}
\end{eqnarray}
Write \ref{ODE} in a matrix form by defining
\begin{itemize}
\item Let $\tilde \chi_n$ be $n\times 1$ vector with $i$th component $\partial^2_{mk}\sigma_i$.
\item Let $\tilde\tau_n$ be $n\times 1$ vector with $i$th component 
$$\Gamma_{mk}
+\Gamma_{\bar\sigma k}\partial_m\bar\sigma_i+\Gamma_{\bar\sigma\bar\sigma}\partial_k\bar\sigma_i\partial_m\bar\sigma_i
+\Gamma_{\bar\sigma m}\partial_k\bar\sigma_i.$$
\end{itemize}
Then \ref{ODE} can be written as
$$(I_n-D_n)\tilde \chi_n=\tilde \tau_n.$$
As we have shown before, $D_n$ is invertible. For any $i\in \mathcal{N}_n$, $|\tau_i|\le B_{\theta,\theta}+2B_{\bar\sigma\theta}C_{\partial\sigma}+B_{\bar\sigma,\bar\sigma}C_{\partial\sigma}^2$, so that $||\tau_n||_\infty=\max_i |\tau_i|$ is uniformly bounded. Therefore,
$$||\tilde x_n||_\infty\le \frac{C_{\tau}}{1-\lambda} $$
and the result follows. $\blacksquare$

%=============================================================================%

\end{appendix}

%=============================================================================%
\newpage
\bibliographystyle{plainnat}
 %---> labelling in appearing order
\bibliography{bibib}    %---> bibliog.bib contains bibtex information

\end{document}